\documentclass[a4paper,twocolumn]{article}

\usepackage[utf8]{inputenc}
\usepackage[T1]{fontenc}
\usepackage{float}
\usepackage{amsmath,amssymb}
\usepackage{mathrsfs}
\usepackage{theorem}
\usepackage{graphicx}
\usepackage{color}
\usepackage{wasysym}
\usepackage{hyperref}

\definecolor{blue}{rgb}{0,0,1}
\definecolor{green}{rgb}{0,0.65,0.5}
\definecolor{verde}{rgb}{0.,.5,0.4}
\definecolor{marron}{rgb}{0.7,0.2,0.1}
\definecolor{red}{rgb}{1,0,0}
\definecolor{vio}{rgb}{0.66,0,1}
\definecolor{ama}{rgb}{1,1,0}
\definecolor{veroscuro}{rgb}{0.3,0.36,0.33}
\newcommand{\bc}{\begin{center}}
\newcommand{\ec}{\end{center}}
\newcommand{\be}{\nopagebreak[3]\begin{equation}}
\newcommand{\ee}{\end{equation}}
\newcommand{\ba}{\nopagebreak[3]\begin{eqnarray}}
\newcommand{\ea}{\end{eqnarray}}

{\theorembodyfont{\sffamily} }
{\theorembodyfont{\sffamily} }
{\theorembodyfont{\sffamily} }

\begin{document}

\title{\bf 
Constructing balanced equations of motion for
particles in general relativistic theories: 
the general case
}

\author{
Emanuel Gallo
and
Osvaldo M. Moreschi
\\
{\rm \small Facultad de Matem\'{a}tica, Astronom\'\i{}a, F\'\i{}sica y Computaci\'{o}n (FaMAF),} \\
{\rm \small Universidad Nacional de C\'{o}rdoba,}\\
{\rm \small Instituto de F\'\i{}sica Enrique Gaviola (IFEG), CONICET,}\\
{\rm \small Ciudad Universitaria,(5000) C\'{o}rdoba, Argentina.}
}

\maketitle

\begin{abstract}

We present a general approach for the formulation of equations of motion for 
compact objects in general relativistic theories. The particle is assumed to be moving
in a geometric background which in turn is asymptotically flat.
Our approach defines a model for particle like objects which emphasizes different
aspects of a set of gravitating objects; by concentrating in the main contributions
coming from back reaction effects due to gravitational radiation.

\end{abstract}

\section{Introduction}\label{sec:intro}

Recently, the first direct detections of gravitational 
waves\cite{Abbott:2016blz,TheLIGOScientific:2016wfe,Abbott:2016nmj,Abbott:2017vtc,
	Abbott:2017oio,TheLIGOScientific:2017qsa,Abbott:2017gyy}
have been announced. 
Thus a new way to observe the Universe and to receive information from it 
has become operative.
In particular, with current detectors we expect to learn more
about astrophysical processes 
involving systems with presence of strong gravitational fields
and high relative velocities among
compact bodies like black holes or neutron stars.
 
Although the detectors were not operating at maximum power and sensitivity
the first detected signal was strong enough to be easily observed.
When in the near future, the detectors acquire the expected 
designed sensitivity
there will be even better defined signals in which the 
fine physical interpretation of the data will become 
an active area of research.
Furthermore, with the higher sensitivity it comes the bigger cosmic
volume that the detectors can explore, and therefore one also
expect to observe a variety of systems with many different
physical situations.
The work for full numerical relativity calculations will probably
have to extend to unexpected initial conditions.
Under this circumstances the need to estimate the most appropriate
initial conditions of the system, will require for 
physical models that can handle strong gravity and back reaction
in the dynamics.

Among the frequently approximate methods used we can mention the 
post-newtonian methods\cite{Will:2005sn,Blanchet:2013haa,Jaranowski:2013lca}, 
the self-force approach\cite{Poisson:2011nh,Wald:2009ue}, 
and models based on 
effective one body approach\cite{Damour08} techniques, etc.
In particular for the case of a binary system of black holes, it is a
good starting model to think the system as consisting of 
point particles whose dynamics is described in a flat background with 
corrections 
due to the perturbations originated by the rest of the components
of the system and the particles itself. 
All these approaches have their advantages and limitations, 
namely validity range, ease of computation, etc.
However in order to arrive to these kind of
models one needs to settle several issues, as for example
what is the notion of particles in the general relativity framework
that one should use. 

Several related questions arise: 
what is the convenient structure that one should assume for the corresponding particle?
in what way one could refer to the related equations of motion?
how should one determine the corresponding equations of motion?

 In reference \cite{Gallo:2011tf} we have presented a new derivation of the equation
of motion of charged particles in a flat background, interacting with an  non-radiating external field,
which balance the electromagnetic radiation.
The new equations of motion is
\begin{equation}\label{eq:balance3m}
(\dot m - \ddot \alpha) v^a + m  \dot v^a 
+ \alpha \ddot v^a
=
e F(B)^a_{\;b }\, v^b 
+
\frac{2}{3} e^2 
 \dot v^b \dot v_b  v^a 
;
\end{equation}
where we are using abstract indices $a,b,...$, and beyond the standard
mass $m$, four velocity $v^a$, charge $e$ and electromagnetic field tensor $F$,
there appears the new degree of freedom $\alpha$, depending on the proper time $\tau$,
a dot over a scalar means derivative with respect to $\tau$ and over a vector means
the covariant derivative in the direction of the vector $v^a$.

This equation can also be decomposed in terms of its projection in the direction of 
the velocity
\begin{equation}\label{eq:mddotalpha}
\dot m - \ddot \alpha
+ \alpha a^2
=
- \frac{2}{3} e^2  a^2
;
\end{equation}
and its orthogonal complement
\begin{equation}\label{eq:balance4m}
m  \dot v^a 
=
e F(B)^a_{\;b }\, v^b 
- \alpha (\ddot v^a - a^2 v^a  )
.
\end{equation}
We have shown that this approach is successful for the derivation of the 
equations of motion, since it agrees with the well known Lorentz-Dirac
dynamics for a particular case.

It is our goal to construct the analog to equation (\ref{eq:balance3m})
for the gravitational case;
where the variation of total momentum per unit time is balanced with
radiation at infinity and some delicate issues related to this problem are solved. 
The present approach for building the equations of motion has in mind that one is concerned with
the first order back reaction effects coming from the fact that accelerated masses
radiate gravitationally.
When dealing with complex systems, if one is interested in higher order effects,
then one should improve the model to the appropriate order.
We here focus on the presentation for the construction of the equations of motion 
that take into account the first order back reaction effects due to gravitational radiation,
within a wide class of general relativistic theories;
previous to the fixing of
a particular field equation.
We do assume however that the energy momentum tensor is divergent free,
and that isolated bodies are well represented by asymptotically flat spacetimes\cite{Moreschi87}.

One may ask why another work should be studied that considers 
the momentum balance as a starting point for the derivation of the equations 
of motion. However, in this article we deal with several aspects that are not 
usually discussed in the literature, and what is more, 
regardless of the final field equations.

Of course referring to particles in the context of general relativity
raises several issues. To begin with, the concept of particle involves
the idea of a very compact object; but when one makes an object smaller
and smaller, one does not end with a point like object but with 
a black hole. Secondly, if one defines particles as objects with
a distributional energy momentum tensor with support on a timelike
world line, one faces the problem of regularity of the spacetime.
In this respect in reference \cite{Geroch87} Geroch and Traschen define
a notion of regular manifolds that admit distributions; and one dimensional
distributions are not allowed in 4-dimensional manifolds.
But one can always consider one dimensional distributions as the sources
of the linear field equation on backgrounds;
and this is the approach we use here.

In particular, we will present here the formal aspect and the problems that could arise in
a model for a compact object, treated as a particle on
an appropriately chosen flat background. The idea one has in mind is to 
apply this construction to a binary system, so that each of the compact
objects will be treated likewise.
The flat background metric by construction will share the same asymptotic
region as the full metric of the spacetime: so that one of the inertial
systems at infinity would be related to this flat global metric.
In the present work, we only use the linear structure of the
undetermined field equations. We will clarify this point below.

We take the opportunity in this article to study all the issues 
that one should tackle in the construction of balanced equations of motion for particles
in general relativistic theories, 
\emph{prior to the choice of field equations and
specific gauges} in the calculation.
The relation that we will make with the asymptotic structure of isolated systems
is very detailed and we have not found similar analysis in the literature.
The main objective in this work is the construction of the general form of a balanced
equations of motion,  for a wide class of field equations, 
which are independent of the gauge one is using.
For this reason, some of the topics that we discuss here,
and
the differences mentioned above, of our model with other approaches, can only
be investigated in detail after the field equations and specific gauges have
been settled.

Since in this work we deal with several issues related to the project of
constructing balanced equations of motion that have been neglected
in the past, we need to go through all these issues in detail;
for this reason we organized this paper as follows.
The basic concepts inherent to a description of a particle, along
with a precise relation between intrinsic angular momentum
and center of mass is presented in section \ref{sec:prel}.
The description of the local
dynamics of the particle is done in section \ref{sec:ldinamics}.
The problematic of asymptotically flat spacetimes at future null infinity,
including the supertranslation problem, the appropriate
description of global physical quantities, and a gauge
invariant treatment of gravitational radiation, 
is presented in section \ref{sec:gaf}.
A detailed discussion of the relation between internal
convenient tetrads and asymptotic convenient tetrads,
along with a suggestion of a new asymptotic flat background
metric associated with the asymptotic center of mass, is
carried out in section \ref{sec:tetra}. 
All these preliminaries are taken into account
in the presentation of our main result which is the 
general form of the balanced equations of motion, appearing
in section \ref{sec:times}.
We conclude with general remarks and comments in the last section.

\section{Preliminaries}\label{sec:prel}

In this section we review some basic concepts in special relativity
that include center of mass, intrinsic angular momentum and distributions; 
with the intention to clarify in detail the fundamental concepts
that we are generalizing to the dynamics of a compact object,
and so prevent future explanatory discussions.

So, since we are warning of the basic level of this section, we advice,
for the reader interested in getting immediately into the subject, to just
jump to the third following section.

\subsection{Particles as isolated objects in \\
special relativity}\label{sec:special-relativity}

The notion of a point like object in a fixed background is normally presented
through an energy momentum tensor which has support only on a timelike world line.

The first case one should understand is of course the flat background case.

Consider the situation in which a distribution of matter
that has support within a world tube of finite size. Later we will take the limit
in which the tube collapses to a timelike line.

Having a flat background we have at our disposal the translational Killing vectors
$K_{\underline{a}}=\frac{\partial}{\partial x^{\underline{a}}}$, 
with $\underline{a}=0,1,2,3$, and also the rotational Killing vectors 
$K_{\underline{ab}}(\xi)=
\eta_{\underline{ac}}(x^{\underline{c}}-\xi^{\underline{c}})\frac{\partial}{\partial x^{\underline{b}}} - 
\eta_{\underline{bc}}(x^{\underline{c}}-\xi^{\underline{c}})\frac{\partial}{\partial x^{\underline{a}}}$;
which are the generator of Lorentz rotations around the point $\xi^{\underline{c}}$.

The matter is represented by the energy-momentum tensor $T_{ab}$. Note that we are using
the notation $a,b,c$ for abstract indices while underlined indices are numeric.
Then, at zero order, the energy-momentum tensor satisfies
that its divergence is zero, namely:
\begin{equation}
 \nabla \cdot T = 0 \; .
\end{equation}

Let $K^a$ denote any of the Killing symmetries, then, 
for each symmetry one has a conserved quantity;
in fact it has been explained elsewhere \cite{Szabados04,Gallo:2011tf} 
how these conserved
quantities can be related to spheres that `surround' the sources. 
Let us define the three form $\mathcal{D}_{abc}(K) = T^d_{\;\;e} \; K^e \epsilon_{abcd}$;
then its exterior derivative is
$d\mathcal{D}_{abcd} = k \nabla_f \left(T^f_{\;\;e} \; K^e \right)\epsilon_{abcd}$, where $k$ is a constant.
This exterior derivative vanishes due to the fact that the divergence of $T$ is zero and $K$ is a Killing
symmetry.
Then if $\Sigma$ is a three dimensional region, such that the world tube goes through its interior, the quantity
\begin{equation}
 Q = \int_\Sigma \mathcal{D}(K) ;
\end{equation}
is conserved; in the sense that if $\Sigma'$ is any other three dimensional region, 
such that the world tube goes through its interior,
then, the integration of $\mathcal{D}$ on $\Sigma'$ also gives $Q$.
Another way of seeing this is the following:
if the boundary of $\Sigma$ is the sphere $S$, then for any
other hypersurface $\Sigma'$ with the same boundary the calculation
of the quantity $Q$ will give the same value;
so that at the end, one associates $Q$ to a given surface $S$.
In this work, we use analogous ideas for the case of gravitating objects,
and we will think of these spheres as residing at future null infinity;
where the total momentum and flux of gravitational radiation are evaluated.

Using these structure one defines $P_{\underline{a}}$ by
\begin{equation}
 P_{\underline{a}} = \int_\Sigma \mathcal{D}(K_{\underline{a}}) ;
\end{equation}
as the components of the conserved total momentum, and also
\begin{equation}
 J_{\underline{ab}}(\xi) = \int_\Sigma \mathcal{D}\big(K_{\underline{ab}}(\xi)\big) ;
\end{equation}
as the components of the conserved total angular momentum.

Let us observe that
\begin{equation}
 J_{\underline{ab}}(\xi_2) - J_{\underline{ab}}(\xi_1) =
\eta_{\underline{ac}}(\xi_1^{\underline{c}}-\xi_2^{\underline{c}})\; P_{\underline{b}} - 
\eta_{\underline{bc}}(\xi_1^{\underline{c}}-\xi_2^{\underline{c}})\; P_{\underline{a}}
;
\end{equation}
or in other words
\begin{equation}
 J^{\underline{ab}}(\xi_2) = J^{\underline{ab}}(\xi_1) +
(\xi_1^{\underline{a}}-\xi_2^{\underline{a}})\; P^{\underline{b}} - 
(\xi_1^{\underline{b}}-\xi_2^{\underline{b}})\; P^{\underline{a}}
.
\end{equation}

It is probably worthwhile to notice that if one acts with Lorentz transformations
on the initial frame 
$e_\text{\underline{a}} \equiv \frac{\partial}{\partial x^\text{\underline{a}}}$,
then the set of quantities $P_{\underline{a}}$ transforms as components
of a co-vector, while the set $J_{\underline{ab}}$ transforms as the components
of an antisymmetric tensor.

Let us emphasize again that $J^{\underline{ab}}(\xi)$ is conserved for any $\xi$.
But one could consider, for example $\xi_1(\tau_1)$ as a timelike world line
with proper time $\tau_1$. 
Therefore, calling $v_1^{\;\underline{a}} = \frac{d\xi^{\underline{a}}}{d\tau_1}$
one has
\begin{equation}
\frac{d}{d\tau_1}  J^{\underline{ab}}(\xi_1) = 
v_1^{\;\underline{b}}\; P^{\underline{a}}
-
v_1^{\;\underline{a}}\; P^{\underline{b}}
.
\end{equation}
Note that if one chooses $v_1$ to be parallel to $P$, then
\begin{equation}
 \frac{d}{d\tau_1}  J^{\underline{ab}}(\xi_1) = 0
.
\end{equation}

Also, let us note that fixing $\xi_2$, one has
\begin{equation}
J^{\underline{ab}}(\xi_1) P_{\underline{b}}
=
 J^{\underline{ab}}(\xi_2) P_{\underline{b}} 
-
(\xi_1^{\underline{a}}-\xi_2^{\underline{a}})\; P^{2} 
+
(\xi_1^{\underline{b}}-\xi_2^{\underline{b}})P_{\underline{b}}\; P^{\underline{a}}
 ;
\end{equation}
so that one can always pick a $\xi_1^{\underline{a}}$ so that 
\begin{equation}\label{eq:jconp}
J^{\underline{ab}}(\xi_1) P_{\underline{b}}
= 0
 .
\end{equation}
Using a rest frame, in which the generator of time translations
is aligned with the total momentum $P^{\underline{a}}$; in other words,
its spacelike components are zero; i.e., $P^i=0$
with $i=1,2,3$; one has $J^{\underline{i}\,0}(\xi_1)=0$ and so
\begin{equation}
 \int_\Sigma (x^{\underline{i}} - \xi_1^{\underline{i}}) T^{00} \epsilon_\Sigma
=
\int_\Sigma (x^{\underline{0}} - \xi_1^{\underline{0}}) T^{i0} \epsilon_\Sigma
=
(x^{\underline{0}} - \xi_1^{\underline{0}}) P^{i}
=0 ;
\end{equation}
and therefore one arrives at the center of mass coordinates
\begin{equation}
 \xi_{\text{cm}}^{\underline{i}} = \frac{1}{M} \int_\Sigma x^{\underline{i}} T^{00} \epsilon_\Sigma
;
\end{equation}
where now $\Sigma$ is an adapted hypersurface $x^0=$constant, $\epsilon_\Sigma$ is the
volume element, and we have used $P^0=M$ in the rest frame.
For this reason condition (\ref{eq:jconp}) is used\cite{Dixon:1970zza} for
defining \emph{center of mass} in special relativity.
Then all other points that are translated parallel to $P$ are also center of mass points;
which means that the center of mass world line has velocity $v = \frac{P}{M}$.

Now, let us consider the case in which the world tube gets smaller until
it collapses to the timelike world line $z$. In this case $x^{\underline{i}}$
is compelled to be evaluated at the world line; and since $P^0=M$,
one has that the center of mass $\xi_{\text{cm}}^{\underline{i}}$
is at the world line $z$.

\subsection{The intrinsic angular momentum}\label{subsec:intrinsic}

The \emph{intrinsic angular momentum} $S_{\underline{a}\underline{b}}$ is defined as the angular momentum
evaluated at the center of mass, namely:
\begin{equation}\label{eq:intrinsictensor1}
S_{\underline{a}\underline{b}} = J_{\underline{a}\underline{b}}(\xi_{\text{cm}}) 
.
\end{equation}

Since any contraction of $S_{\underline{a}\underline{b}}$ with $P^{\underline{a}}$
gives zero, there are only three degrees of freedom involved in the intrinsic angular
momentum; so that, similarly as the magnetic field is defined in terms of the
electromagnetic tensor and the volume element,
one defines the spin vector from
\begin{equation}\label{eq:apulilubanski}
 S_{\underline{a}} = \frac{1}{2} 
\epsilon_{\underline{a}\underline{b}\underline{c}\underline{d}}
J^{\underline{b}\underline{c}}(\text{any }\xi) P^{\underline{d}}
;
\end{equation}
which by construction is orthogonal to $P^{\underline{a}}$, and so it only 
involves three degrees of freedom.
This is also known as the Pauli-Lubanski pseudovector.

In (\ref{eq:apulilubanski}) we have emphasized the fact that the spin vector
can be calculated with respect to any choice of reference center $\xi$;
since the contraction with the momentum vector eliminates this dependence.

Then one can think in reconstructing the complete angular momentum tensor
from the spin vector.
Let us note that
\begin{equation}
\begin{split}
\epsilon^{\underline{a}\underline{b}\underline{c}\underline{d}}
S_{\underline{c}} P_{\underline{d}}
=&
\frac{1}{2} 
\epsilon^{\underline{a}\underline{b}\underline{c}\underline{d}}
P_{\underline{d}}
\epsilon_{\underline{c}\underline{e}\underline{f}\underline{g}}
J^{\underline{e}\underline{f}} P^{\underline{g}} \\
=&
-
\left( 
M^2  J^{\underline{a}\underline{b}} 
+
P^{\underline{a}}  J^{\underline{b}\underline{d}} P_{\underline{d}}
-
P^{\underline{b}}  J^{\underline{a}\underline{d}} P_{\underline{d}}
\right)
\end{split}
\end{equation}

One can see then that the intrinsic angular momentum tensor, in terms of the spin vector, 
is given by
\begin{equation}\label{eq:intrinsictensor2}
 S^{\underline{a}\underline{b}} =
\frac{1}{M^2}
\epsilon^{\underline{a}\underline{b}\underline{c}\underline{d}}
P_{\underline{c}} S_{\underline{d}}  .
\end{equation}

Although we have previously introduced first the concept of center of mass and
then the one of intrinsic angular momentum;
this last equation allows us to define the center of mass,
in terms of the intrinsic angular momentum,
in the following way:
\begin{itemize}
 \item 
Let us first define the intrinsic spin vector $S^{\underline{a}}$ by:
\begin{equation}\tag{\ref{eq:apulilubanski}}
 S_{\underline{a}} = \frac{1}{2} 
\epsilon_{\underline{a}\underline{b}\underline{c}\underline{d}}
J^{\underline{b}\underline{c}}(\text{any }\xi) P^{\underline{d}}
.
\end{equation}

\item Then let us define the intrinsic angular momentum tensor
$S^{\underline{a}\underline{b}}$ from:
\begin{equation}\tag{\ref{eq:intrinsictensor2}}
 S^{\underline{a}\underline{b}} =
\frac{1}{M^2}
\epsilon^{\underline{a}\underline{b}\underline{c}\underline{d}}
P_{\underline{c}} S_{\underline{d}}  .
\end{equation}
Let us emphasize that this definition only employs intrinsic properties
of the system, i.e.; it does not recourse to any additional structure.

\item Finally the center of mass points are defined as those that satisfy:
\begin{equation}\tag{\ref{eq:intrinsictensor1}}
S_{\underline{a}\underline{b}} = J_{\underline{a}\underline{b}}(\xi_{\text{cm}}) 
.
\end{equation}

\end{itemize}

We conclude then that there is a one to one relation between the concept of
\emph{center of mass} and \emph{intrinsic angular momentum}.

Then the total angular momentum with respect to an arbitrary center $\xi$
can be expressed in terms of the intrinsic angular momentum and
the center of mass by:
\begin{equation}\label{eq:tot-int-cent}
 J^{\underline{a}\underline{b}}(\xi) = S^{\underline{a}\underline{b}}
+ P^{\underline{a}} \, (\xi^{\underline{b}}-\xi_\text{cm}^{\underline{b}})
- P^{\underline{b}} \, (\xi^{\underline{a}}-\xi_\text{cm}^{\underline{a}})
\, .
\end{equation}

One can see that the spin vector $S^a$ is also conserved along the world line.

It is also clear from this analysis that \emph{elementary particles} can have momentum and spin
but not dipole; as it has been frequently assumed in the literature.
By dipole we precisely mean $d^a = S^{ab} P_b$.

We have not considered here the case of electrically charged particles since the 
energy momentum tensor we have considered are of finite space support.

Of course one could consider \emph{composite particles} which have extra structure,
beyond momentum and spin; and then one is free to include the extra dynamics
for the extra degrees of freedom, compatible with the conservation of the energy
momentum tensor.

\subsection{Particles as compact objects in 
curved spacetimes} 

One can use the concept of particle when the dimension of the object and its
structure is such that only the monopole nature is enough to describe its
dynamics. In the language of Mathisson\cite{Mathisson37,Mathisson40}, the 
\emph{dynamical skeleton} is described by a monopole.

Let us assume that in a neighborhood of the timelike curve $C$
we have at our disposal coordinates $x^{\underline{a}}$, so that the curve
is given by $x^{\underline{a}}=z^{\underline{a}}(\tau)$, with
$\frac{dz^0}{d\tau}>0$.
At each point along the curve $C$ we can express a 3-dimensional 
delta function $\delta^{(3)}$, in terms of local coordinates,
with support on $C$ by
\begin{equation}
\begin{split}
 \delta^{(3)}&
\big( (x^1-z^1(\tau)) ,(x^2-z^2(\tau)) ,(x^3-z^3(\tau)) \big) \\
=&
\frac{1}{\frac{dz^0}{d\tau}\sqrt{- \det{g}}}
 \delta(x^1-z^1(\tau)) \delta(x^2-z^2(\tau)) \delta(x^3-z^3(\tau)) 
;
\end{split}
\end{equation}
where $\det{g}$ is the determinant of the four metric.

Then, an energy momentum tensor for a monopole particle
with support on a timelike worldline $C$
is represented\cite{Wald:2009ue,Poisson:2011nh} by
\begin{equation}\label{eq:tab1}
\begin{split}
&T_{ab}(x^0=z^0(\tau),x^1,x^2,x^3) =
\, M\; v_{a}(\tau)\; v_{b}(\tau) \\
& 
\frac{1}{\frac{dz^0}{d\tau}\sqrt{- \det{g}}}
 \delta(x^1-z^1(\tau)) \delta(x^2-z^2(\tau)) \delta(x^3-z^3(\tau)) ;
.
\end{split}
\end{equation}
Below we will specify further the expression that is used in this work.

In the following sections we present the construction of the model.

\section{Local dynamics at zero order in back reaction}\label{sec:ldinamics}

As mentioned above, one of the objectives of the present work is to discuss several
general issues that must be taken into account in the construction of the equations
of motion that can be applied to a binary system, in the dynamical situation
in which the back reaction effects due to gravitational radiation are important.
The model is described in terms of an asymptotically flat spacetime at
future null infinity.
For this reason we concentrate in one particle, to which we assign the 
label $A$. The rest of the system, which it could be another particle is assigned
label $B$.

It is our intention to construct a model for a binary system 
that takes into account the leading order contributions to the equations 
of motion.
This is enough to force us to discuss a variety of issue that appear
in this type of problems.
In order to obtain a high degree of precision in the predicting
power of the equations of motion, one should extend this work
to higher orders.
In separate works we will present what we call first order
equations of motion, in the harmonic gauge, and second
order equations of motion in the null gauge.
This instead is the general treatment of the first order
version of the balanced equations of motion.

Let the metric $\mathsf{g}$ be the exact metric that corresponds to an isolated 
binary 	system of compact objects. Then there are infinite ways in which
it can be decomposed in a form:
\begin{equation}
\mathsf{g}  = \eta + \mathsf{h}_A + \mathsf{h}_B + \mathsf{h}_{AB} ;
\end{equation} 
where $\eta$ is a flat metric,
$ \mathsf{h}_A$ is proportional to a parameter $M_A$, that one can think is some kind
of measure of the mass of system $A$,
similarly $ \mathsf{h}_B$ is proportional to a parameter $M_B$,
and $ \mathsf{h}_{AB}$ is proportional to both parameters.

Since we will deal with back reactions for each body, we need to
single out, at future null infinity, the gravitational radiation 
corresponding to each object. Therefore, we need, at this stage
of the dynamics, to be able to distinguish terms in the geometry
that correspond to each of the binary components.

To study the gravitational radiation emitted by the motion of
particle $A$, 
we will model the asymptotic structure of a sub-metric 
\begin{equation}\label{eq:gA}
g_A = \eta + h_A ;
\end{equation}
in order to calculate the reaction due to gravitational radiation.

We adjust $g_A$ to model a compact object, which dynamics
is affected by the existence of system $B$, and therefore
by the geometry determined by a sub-metric
\begin{equation}\label{eq:gB}
g_B = \eta + h_B .
\end{equation}
The appropriate choice of the flat metric $\eta$ 
was discussed in article \cite{Gallo:2016hpy};
the basic idea is that it should be related to a local notion
of center of mass frame.

For the back reaction calculations, 
we need to capture the
leading order behavior of the metric;
although taking 
$h_A$ to agree in the linear regime with $\mathsf{h}_A$ 
and also $h_B$ to agree in the leading order with $\mathsf{h}_B$
is enough for
first order back reaction, gravitational radiation effects.

Although in this work we will mainly use the geometries (\ref{eq:gA})
and (\ref{eq:gB}), one can think that the model represents 
the combined system by
a leading order approximation of the spacetime metric of the form
\begin{equation}\label{eq:gAB}
g = \eta + h_A + h_B 
;
\end{equation}
which is also assumed to be asymptotically flat at future null
infinity; so that one can discuss the effects of gravitational radiation,
and the metric can also be expressed as (\ref{eq:asymp1}).
For the purpose of accuracy in the dynamical calculations,
we need $h_A$ and $h_B$ to represent as close as possible
the exact geometry of the binary system;
and therefore one should pursue this approach to higher orders;
trying to get as close as possible to the exact
representation of the metric.

Although we will be studying the dynamics of system $A$, to simplify the 
notation we will avoid using a subindex $A$, whenever possible.

The expression for the momentum of a particle can be deduced from the conservation law for the energy-momentum
tensor.

The limit in which one considers particle $A$ to be `small' enough so that the 
radiation effects can be neglected, gives rise to the zero order in $\lambda$
approximation; where $\lambda$ is a parameter associated to the strength of
gravitational radiation, that will be clarified below.

The zero order in $\lambda$ equations of motion for particle $A$ comes from
the zero order conservation equation for the energy momentum tensor of this particle;
namely
{
\begin{equation}
 \nabla_{(B)a}\; T_{(A)}^{\quad ab} = 0 .
\end{equation}
}

This zero order equations of motion is considered in terms of an
energy-momentum which is a distribution on the background geometry $g_{(B)}$.
Then, at this order, for the energy momentum tensor of a monopole body $A$, one must consider
\begin{equation}
\begin{split}
&T_{(A) ab}(x^0=z^0(\tau),x^1,x^2,x^3) =
\, M_A\; v_{(A) a}(\tau)\; v_{(A) b}(\tau) \\
&
\frac{1}{\frac{dz^0}{d\tau}\sqrt{- \det{g}_{(B)}}}
 \delta(x^1-z^1(\tau)) \delta(x^2-z^2(\tau)) \delta(x^3-z^3(\tau)) 
;
\end{split}
\end{equation}
where 
$\det g_{(B)}$ is the determinant of the metric $g_{(B)}$
and $(x^1,x^2,x^3)$ are local spacelike coordinates.

Consider a small world tube around the world line of the particle.
Let $e_{\underline{c} }^{\;\; b}$ be a frame basis
(an orthonormal frame with respect to metric $g_{(B)}$);
where, as before, 
we are distinguishing between the abstract indices $(a,b,c,..)$ and the numeric frame indices
$(\underline{a},\underline{b},\underline{c},...)$ and let $\theta^{\underline{c}}_{\;\;b}$
be its coframe.
Then one can integrate, in the interior of the world tube, the expression
\begin{equation}\label{eq:int-div}
\begin{split}
 0 =& \int \nabla_{(B)}^{\quad a}\left( T_{(A) ab} \right)\; e_{\underline{c} }^{\; b} dV^4 \\
=&
\int \nabla_{(B)}^{\quad a}\left( T_{(A) ab}\; e_{\underline{c} }^{\; b} \right) dV^4 
-
\int  T_{(A) ab}\; \nabla_{(B)}^{\quad a}\left( e_{\underline{c} }^{\; b} \right) dV^4 \\
=&
\int  T_{(A) ab}\; e_{\underline{c} }^{\; b}\; \mathtt{t}_2^a \; dV_2^3
-
\int  T_{(A) ab}\; e_{\underline{c} }^{\; b}\;  \mathtt{t}_1^a \; dV_1^3-\\
&
\int  T_{(A) \;\;b}^{\quad \underline{a}}\; 
\Gamma_{(B)\underline{a} \; \underline{c}}^{\quad \;\; \underline{b}} \; e_{\underline{b} }^{\; b}  dV^4
;
\end{split}
\end{equation}
where $\mathtt{t}_2^a$ is the vector gradient at 
the future boundary, $\mathtt{t}_1^a$ is the corresponding vector at 
the past boundary, of an adapted time coordinate $t=x^0$, namely $\mathtt{t}^a = g_{(B)}^{\quad ab} \, dt_b$, and $\Gamma_{(B)\underline{a} \; \underline{c}}^{\quad \;\; \underline{b}}$ is the torsion free metric connection of $g_{(B)}$.

We can always reparameterize the
local time $t$, to adapt it to this geometric construction, so that the future boundary
coincide with $t +d\tau$ and the past boundary coincide with $t$; where $d\tau$
is the increment in proper time along the curve $C$, between the two intersection
points of the curve with the future and past boundary hypersurfaces of the world tube.
Then carrying out the integration on the right hand side, one obtains
\begin{equation}
\begin{split}
 & M_A \, g_{(B)}(v_{(A)}, e_{\underline{c} }) \, g_{(B)}(v_{(A)}, \mathtt{t} ) \big|_{t +d\tau} \\
-& M_A \, g_{(B)}(v_{(A)}, e_{\underline{c} }) \, g_{(B)}(v_{(A)}, \mathtt{t} ) \big|_{t } \\
-&
 M_A \, v_{(A)}^{\underline{a}} \, v_{(A)}^{\underline{d}} \, g_{(B)b\underline{d}} \,
\Gamma_{(B)\underline{a} \; \underline{c}}^{\quad \;\; \underline{b}} \; e_{\underline{b} }^{\; b}
d\tau
;
\end{split}
\end{equation}
where all the calculation is done with respect to the background metric $g_{(B)}$.
Let us note that, from the choice of $t$, we have
$g_{(B)}(v_{(A)}, \mathtt{t} )= v_{(A)}(t) =1$.

Then, the local notion of momentum is obtained, from
\begin{equation}
 P_{(A)\underline{c}} = {M_A} g_{(B)}(v_{(A)}, e_{\underline{c} })
;
\end{equation}
so that one has
\begin{equation}\label{eq:papa}
0=
\left.   P_{(A)\underline{c}}\right|_{\tau +d\tau}
-
\left. P_{(A)\underline{c}}\right|_{\tau }
-
M_A v_{(A)}^{\underline{a}} v_{(A)}^{\underline{d}} g_{(B)b\underline{d}}
\Gamma_{(B)\underline{a} \; \underline{c}}^{\quad \;\; \underline{b}} \; e_{\underline{b} }^{\; b}
{
\; d\tau
}
.
\end{equation}

Dividing by $d\tau$ and taking the limit to zero, the first two terms constitute the total derivative
\begin{equation}
\begin{split}
 \frac{d}{d\tau}  P_{(A)\underline{c}} 
=&
v_{(A)}^{\quad e} \nabla_{(B)e}  P_{(A)\underline{c}} \\
=&
 M_A \; g_{(B)ab}
\left(
v_{(A)}^{\quad e} \nabla_{(B)e}(v_{(A)}^{\quad a}) e_{\underline{c} }^{\; b}\right.\\
&\left.+
v_{(A)}^{\quad a} v_{(A)}^{\underline{d}} 
\Gamma_{(B)\underline{d} \; \underline{c}}^{\quad \;\; \underline{b}} \; e_{\underline{b} }^{\; b}
\right)
;
\end{split}
\end{equation}
where in the first equality we are using that the vector $v_{(A)}^{\quad e}$
has unit modulus with respect to $g_{(B)}$.
Then substracting the last term of (\ref{eq:papa}) one obtains, from (\ref{eq:int-div}),
at zero order in back reaction due to gravitational radiation, the well known equation
\begin{equation}\label{eq:dpdtau}
\begin{split}
M_A \, g_{(B)ab}
\left(
v_{(A)}^{\quad e} \nabla_{(B)e}(v_{(A)}^{\quad a}) e_{\underline{c} }^{\; b}
\right)
= 0
.
\end{split}
\end{equation}

In what follows we will study the corrections to this equation
coming from back reaction effects due to gravitational radiation.

In order to make a connection with the asymptotic discussion we identify the 
basis field vectors $e_{\underline{c} }^{\;\; b}$ with four translational Killing
vectors of the background metric in the asymptotic region.

\section{General asymptotically flat spacetimes}\label{sec:gaf}

\subsection{Particles as isolated objects in \\
 curved spacetimes}
When considering a compact object as isolated, in the framework of isolated spacetimes,
one realizes that in general one can ascribe a flat background to the spacetime.
More concretely, in the asymptotic region one can always write the metric as
\begin{equation}\label{eq:asymp1}
 g = \eta_{\text{asy}} + h_{\text{asy}} 
;
\end{equation}
where $\eta_{\text{asy}}$ is a flat metric associated to inertial frames in the asymptotic
region
 and $h_{\text{asy}}$ the tensor where all the physical information is encoded.

In reference \cite{Moreschi86} we have indicated how to construct a
family of inertial Bondi frames in a neighborhood of future null infinity.
They include the determination of a null tetrad and associated null coordinate
system in the asymptotic region.
For each choice of inertial frame, the metric adopts a decomposition
of the form (\ref{eq:asymp1}).
But, the warning is that there are as many inertial frames as there are
proper supertranslation generators of the asymptotic BMS\cite{Sachs62}
symmetry group.
Therefore, there are as many flat metrics $\eta_{\text{asy}}$ as there are 
proper supertranslation generators.
This is the reason why this asymptotic region is not asymptotically
Minkowskian; since there are too many flat background geometries.
By asymptotically Minkowskian we mean a spacetime which in the asymptotic
region is decomposed in terms of a unique flat metric.

Also, one knows that the difficulties in finding appropriate rest frames
comes from the existence of gravitational radiation\cite{Moreschi88,Moreschi98,Dain00'}.
Although in the past we have found a way to select rest frames based on the notion
of center of mass and intrinsic angular momentum\cite{Moreschi04,Gallo:2014jda}.
In other words, for each point at future null infinity, we have a way to single out
a unique decomposition of the metric in the form (\ref{eq:asymp1}), with an
appropriately selected flat background.

This complicated structure of rest frames in the asymptotic region
	and its subtle relation with corresponding symmetries of flat backgrounds
	in the interior has normally been neglected in works dealing with
	equations of motion of particles in relativistic theories;
	on the contrary we choose to take into account these effects
	at the beginning for the discussion of back reaction on the motion
	of compact objects. 

\subsection{The leading order behaviour of an adapted null tetrad}

Let now $u$ denotes a null hypersurface that contains future directed null geodesics
that reach future null infinity.

One can express the asymptotic geometry in terms of a complex null tetrad
$\left( \ell ^{a},m^{a},\overline{m}^{a},n^{a}\right) $ with the properties:
\begin{equation}\label{eq:produc}
g_{ab}\;\ell ^{a}\;n^{b}=-g_{ab}\;m^{a}\;\overline{m}^{b}=1 ,
\end{equation}
and all other possible scalar products being zero; so that the metric can be
expressed by
\begin{equation}
g_{ab}=\ell _{a}\;n_{b}+n_{a}\;\ell _{b}-m_{a}\;\overline{m}_{b}-\overline{m}%
_{a}\;m_{b} \, .
\end{equation}

Here, and whenever dealing with null tetrads we will make use of the GHP\cite{Geroch73}
notation.

Using the null polar coordinate system 
$(x^{0},x^{1},x^{2},x^{3})=\left( u,r,,x^{2},x^{3})\right)$
one can express the null tetrad as:
\begin{equation}
\ell _{a}=\left( du\right)_{a} 
\label{uno}
\end{equation}
\begin{equation}
\ell ^{a}=\left( \frac{\partial }{\partial \,r}\right) ^{a} 
\label{dos}
\end{equation}
\begin{equation}
m^{a}=\xi ^{i}\left( \frac{\partial }{\partial x^{i}}\right) ^{a} 
\label{eq:vecm}
\end{equation}
\begin{equation}
\overline{m}^{a}=\overline{\xi}^{i}\left( \frac{\partial }{\partial x^{i}}\right) ^{a} 
\label{tres}
\end{equation}
\begin{equation}\label{eq:vecn}
n^{a}=\,\left(\frac{\partial}{\partial \,u} \right)^{a}
+ \,U\,\left( \frac{\partial }{\partial \,r}\right)^{a}
+ X^{i}\,\left(\frac{\partial }{\partial \,x^{i}}\right)^{a} 
\end{equation}
with $i=2,3$.

\subsection{The total momentum}
Given any section $S$ at future null infinity, the total momentum
{of a generic spacetime}, in terms of an
inertial (Bondi) frame\cite{Moreschi86}, is normally given\cite{Penrose63} by
\begin{equation}\label{eq:totalp}
\mathcal{P}^{\nu} =  - \frac{1}{4 \pi} \int_S {\hat l}_0^{\nu} 
(\Psi_2^0 + \sigma_0 \dot {\bar \sigma}_0) dS^2 ,
\end{equation}
where
the auxiliary null vector 
$\hat l_0(x^2,x^3)$, is defined
in terms of the angular coordinates $(x^2,x^3)$,
by
\begin{equation}  \label{eq:bflhat}
\begin{split}
\hat l_0^\nu(x^2,x^3)
\equiv &
 \biggl(1,\sin(\theta) \cos(\phi), \sin(\theta) \sin(\phi), \cos(\theta) \biggr)\\
=&\left( 1,\frac{\zeta +\bar \zeta }{
1+\zeta \bar{\zeta }},\frac{\zeta -\bar{\zeta }}{i(1+\zeta
\bar{\zeta )}},\frac{\zeta \bar{\zeta }-1}{1+\zeta \bar{\zeta
}}\right);
\end{split}
\end{equation}
where $\mu, \nu, \cdots =0,1,2,3$,
and we are using either the standard sphere angular coordinates 
$(\theta,\phi)$ or the complex angular coordinates $\zeta =\frac{\hat x^2 + i \hat x^3}{2}$,
where $(\zeta,\bar\zeta)$ are complex stereographic coordinates
of the sphere; which are related to the standard coordinates by
$\zeta = e^{i \phi} \cot(\frac{\theta}{2})$;
and here a dot means $\frac{\partial}{\partial \tilde u}$; i.e. the partial derivative with respect
to the inertial asymptotic time $\tilde u$, $\Psi_2^0$ is a component of the Weyl tensor in the 
GHP notation and $\sigma_0$ is the leading order behavior of the spin
coefficient $\sigma$ in terms of the asymptotic coordinate $\hat r$.
So the set of intrinsic inertial coordinates 
at future null infinity
are $(\tilde u,\theta,\phi)$ or
$(\tilde u,\zeta,\bar\zeta)$.

\subsection{The total momentum and flux from charge integrals}
The total momentum for the monopole particle can be calculated using the
charge integral of the Riemann tensor technique.
In this subsection we will use the notation of reference \cite{Moreschi04}.

Let us note
that for any 2-form $w$ defined on a sphere $S$, one can define the charge integral\cite{Moreschi04}
obtaining
\begin{equation}
  \label{eq:chargegeneral}
\begin{split}
  Q_S(w) = 4 \int &\left[
- \tilde w_2 (\Psi_1 - \Phi_{10}) + \right.
 2 \tilde w_1 (\Psi_2 - \Phi_{11} - \Lambda ) \\
& \left. - \tilde w_0 (\Psi_3 - \Phi_{21})
\right] dS^2 + {\tt c.c.}
\end{split}
\end{equation}
where $(\tilde w_0, \tilde w_1, \tilde w_2)$ are the components of $w_{AB}$ with
respect to the dyad adapted to $S$, namely for which the vectors $m$ and $\bar m$ 
are tangent to $S$ and $\ell$ is outgoing,
and $dS^2$ is the area element induced from the spacetime metric. 
If $w$ has a regular extension to future null infinity, denoted by $\mathscr{I}^+$,
we can take the limit of $S$ to a section of $\mathscr{I}^+$ obtaining
\begin{equation}
  \label{eq:chargescri}
\begin{split}
    Q_S(w) = 4 \int &\left[
-  w_2 \Psi_1^0  + 
  2  w_1 \Psi_2^0 
-  w_0 \Psi_3^0 
\right] dS^2 \\
& + {\tt c.c.}
\end{split}
\end{equation}
where the upper-script $0$ denotes the leading order term expansion of the 
Weyl spinor components. 
The charge integral at $\mathscr{I}^+$ turns out to be independent of
the Ricci tensor since as we have shown in \cite{Moreschi87} for any
asymptotically flat spacetime, the components of the Ricci tensor go to zero 
faster than the accompanying Weyl components in the above terms. 

The motivation for these charges, is that they reproduced the known charges in
the case of linearized Einstein's field equations. In the following, we assume that independent
of the exact nature of the final field equations, they reduce in the linear regime to the same
equations as those obtained starting from Einstein's equations.
Using the equations of reference \cite{Geroch73} one can obtain
expressions  for the asymptotic fields with respect to an inertial
reference frame:
the radiation field is given by
\begin{equation}
 \Psi_3 = \frac{\eth_0 \sigma'_0}{r^2} + \mathcal{O}(\frac{1}{r^3}) ,
\end{equation}
while
the relation between the leading order behavior of the shear and primed shear
is
\begin{equation}
 \dot \sigma_0 = - \bar \sigma'_0 .
\end{equation}
The time derivative of the leading order behavior of $\Psi_2$ is
\begin{equation}
 \dot \Psi_2^0 = \eth_0 \Psi_3^0 + \sigma_0 \Psi_4 ,
\end{equation}
and the leading order behavior of the radiation component is
\begin{equation}\label{eq:radiationPsi4}
 \Psi_4^0 = \dot \sigma'_0 .
\end{equation}

In the standard presentation, in terms of an inertial frame,
 one has 
$\Psi_3^0 = -\eth \dot{\bar \sigma}_0$, and using equation (38) of \cite{Moreschi04}
one can write the charge integral  as
\begin{equation}
  \label{eq:chargescri2}
\begin{split}
    Q_S(w) = 4 \int & \left[
-  w_2 \Psi_1^0  + 
  2  w_1 (\Psi_2^0 + \sigma_0 \dot{\bar \sigma}_0)
\right] dS^2 \\
&+ {\tt c.c.}
\end{split}
\end{equation}
We would rather like to emphasize the natural appearance of the $\sigma'_0$
field in these equations, instead of the normally used $\sigma_0$; so that we express:
\begin{equation}
  \label{eq:chargescri2-b}
\begin{split}
    Q_S(w) = 4 \int & \left[
-  w_2 \Psi_1^0  + 
  2  w_1 (\Psi_2^0 - \sigma_0 \sigma'_0)
\right] dS^2 \\
&+ {\tt c.c.}
\end{split}
\end{equation}
In the calculation for the flux of the momentum one takes $w_2=0$,
and obtains:
\begin{equation}
\begin{split}
 \dot Q_S(w) 
=& 4 \int  \left[
2  w_1 (\dot\Psi_2^0 - \dot\sigma_0 \sigma'_0
- \sigma_0 \dot\sigma'_0)
\right] dS^2 + {\tt c.c.} \\
=& 4 \int  \left[
2  w_1 (\eth_0 \Psi_3^0 + \sigma_0 \Psi_4 - \dot\sigma_0 \sigma'_0
- \sigma_0 \dot\sigma'_0)
\right] dS^2 + {\tt c.c.} \\
=& 4 \int  \left[
2  w_1 ( \bar\sigma'_0 \sigma'_0 )
\right] dS^2 + {\tt c.c.} 
\end{split}
\end{equation}
where the time derivative is with respect of an inertial time.
We conclude that the natural way to encode the radiation flux of momentum
is in terms of the $(\sigma'_0 \bar \sigma'_0)$ factor.

The final form of this radiation field $\sigma'_0$ will depend on the particular gauge
one uses for the calculation; however we advance here that this field
is gauge invariant in the inertial harmonic frame discussion,
as we will show in a separate article.
Instead, the field $\sigma_0$ depends on natural new 
freedoms that appear in the harmonic gauge treatment.

It follows that
the time variation of the total momentum, {of a generic spacetime,} 
can be expressed by
\begin{equation}\label{eq:bondibalance}
\dot{\mathcal{P}}^{\mu} =  - \frac{1}{4 \pi } \int_S 
{\hat l_0}^{\mu} \sigma'_0 {\bar \sigma}'_0 dS^2 
\equiv -\mathcal{F}^{\mu}
;
\end{equation}
that is, $\mathcal{F}^{\mu}$ is the total momentum flux.

Since in the dynamics of a particle one normally makes use of a notion of proper time;
it is natural to express the dynamical equations in terms of this frame independent
time parameter.
This proper time will be related with a corresponding time variable at future
null infinity. 
Let $u=$constant represent a general time coordinate and set of sections
such that
\begin{equation}\label{eq:dutildedu}
 \frac{\partial \tilde u}{\partial u} 
= \tilde V(u,\zeta,\bar\zeta) > 0 ,
\end{equation}
is the time derivative of the inertial time $\tilde u$ 
with respect to the non-inertial time $u$;
so that we can use the inertial coordinates or the non-inertial coordinates
$(u,\zeta,\bar\zeta)$.
This means that given the initial section $S$ 
one can define a new consecutive section $S'$
so that both sections are defined by the condition $u=$constant.

Let us remark that the total momentum can be expressed in terms
of the supermomentum $M$
\begin{equation}
\Psi_M = \Psi_2^0 - \sigma_0 {\sigma}'_0 + \eth^2\bar{\sigma }_0
, 
\end{equation}
that we have presented elsewhere\cite{Moreschi88};
so that the momentum is:
\begin{equation}\label{eq:totalp2}
\mathcal{P}^{\mu} =  - \frac{1}{4 \pi} \int_S {\hat l_0}^{\mu} \,
\Psi_M \, dS^2 ;
\end{equation}
where $S$ is a general section at future null infinity.
An important property of the supermomentum $M$ is the transformation
	law under BMS transformations, namely\cite{Moreschi98}
\begin{equation}
\overset{\AC}{\Psi}_M 
= \frac{1}{K^3} \big(\Psi_M -  \bar\eth^2 \eth^2 \gamma \big)
;
\end{equation}	
where $K$ represents here the Lorentz factor, and $\gamma$ the supertranslation.
From this it can be easily 	deduced that, a supertranslation does not
affect the calculation of the total momentum.
Alternatively, if one uses a section that does not coincide with a section
$\tilde u=$constant, then one can still use expression (\ref{eq:totalp2})
to calculate the momentum.

The time variation of the total momentum with respect to the inertial time 
can be expressed simply
in terms of the time variation of this supermomentum, since
\begin{equation}
  \label{eq:flux}
  \dot\Psi_M = \sigma'_0 \, {\bar \sigma}'_0 .
\end{equation}

Then when evaluating the time variation of the momentum with respect to the 
non-inertial time variable $u$, one needs to consider
\begin{equation}
 \frac{\partial \Psi_M}{\partial u}
=
\tilde V \frac{\partial \Psi_M}{\partial \tilde u}
=
\tilde V \sigma'_0 \, {\bar \sigma}'_0 ;
\end{equation}
so that
\begin{equation}\label{eq:bondibalanceV}
\frac{d\mathcal{P}^{\mu} }{du}
 =  - \frac{1}{4 \pi } \int_S 
{\hat l_0}^{\mu} \tilde V \, \sigma'_0 \, {\bar \sigma}'_0 \, dS^2 
= -\mathcal{F}_V^{\mu}
;
\end{equation}
where now $\mathcal{F}_V^{\mu}$ is the instantaneous momentum flux
with respect to the dynamical time $u$, and $S$ is defined by the
$u=$constant condition.

This means that even when considering non-inertial times, one only
needs to calculate the inertial primed shear to evaluate the flux of gravitational
radiation.

\subsection{Expansion of a radiating spacetime in terms of the strength of
the radiation {\rm (The $\lambda$ expansion)}}

One important aspect of the gravitational field is that for many interesting
astrophysical systems the gravitational energy radiated is weak.
Also most of the difficulties that arise at future null infinity 
in the construction of 
appropriate rest reference frames\cite{Moreschi98} 
come from the presence of
gravitational radiation.
It is therefore tempting to expand the structure equations of an asymptotically
flat spacetime in terms of the strength of the gravitational radiation.

We will use the smallness parameter $\lambda$ and assign to the momentum 
flux the order
$\mathcal{O}(\lambda^2)$.

Every field at $\mathscr{I}^+$ can be expanded as a series with different powers of 
$\lambda$; that is for a field $f(u,\zeta,\bar\zeta)$ we will write
\begin{equation*}
  f(u,\zeta,\bar\zeta) =   f_0(\zeta,\bar\zeta) + \lambda \, f_1(u,\zeta,\bar\zeta) 
                              +   \lambda^2 \, f_2(u,\zeta,\bar\zeta) + O(\lambda^3) .
\end{equation*}
The idea is to carry out this expansion on the `center of mass frame' defined in
terms of `nice sections'\cite{Moreschi88}.

For a general asymptotically flat spacetime\cite{Moreschi87} we 
will define the order $\lambda$ by
asigning $\sigma'_0$ to be of order $\lambda$, that is:
\begin{equation}
 \sigma'_0 \equiv \mathcal{O}(\lambda) ;
\end{equation}
where $\sigma'_0$ is the leading order behavior of the GHP spin coefficient $\sigma'$,
in an asymptotic null inertial frame.

\section{Relation between null tetrads of the flat and curved geometries}\label{sec:tetra}

In this section we will study the relation of the different
tetrads that appear
when one considers the decomposition of the metric $g$ with respect to diverse
backgrounds.

\subsection{The general case}\label{subsec:general}
For each decomposition of the metric of the form
\begin{equation}\label{eq:getamash}
g_{ab} = \eta_{ab} + h_{ab} ;
\end{equation}
we have at our disposal two families of null tetrads; one for each
of the metrics.

Let us call $e_{1,2,3,4}=(l,n,m,\bar m)$ a null tetrad of $g$
and $\mathbf e_{1,2,3,4}=(\mathbf{l},\mathbf{n},\mathbf{m},\bar{\mathbf{m}})$ a null tetrad of $\eta$.
Since both metrics have the same asymptotic future null infinity
we can take those tetrads so that they agree asymptotically.
But one should have in mind that the flat metric $\eta$ that is used in the 
interior of the spacetime, need not agree with an asymptotic inertial flat metric
at future null infinity.
Let us use $G$ for the gravitational constant;
then, $h$ is order $\mathcal{O}(G)$ and one can express
\begin{equation}\label{eq:tetrad-relations}
 e_{\underline{a}} = \mathbf{e}_{\underline{a}} 
+ G \, d_{\underline{a}}^{\,\;\underline{b}} \; \mathbf{e}_{\underline{b}} . 
\end{equation}

Then, taking all the contractions with the metric, up to second order,
 one obtains
\begin{equation}
\begin{split}
g(e_{\underline{a}}, e_{\underline{b}}) =& \eta_{\underline{a} \underline{b}} \\
=& \eta_{\underline{a} \underline{b}} 
{
+ G \, d_{\underline{a}}^{\,\;\underline{c}} \, \eta_{\underline{b} \underline{c}}
+ G \, d_{\underline{b}}^{\,\;\underline{c}} \, \eta_{\underline{a} \underline{c}}
}
+ G^2 \, d_{\underline{a}}^{\,\;\underline{c}} d_{\underline{b}}^{\,\;\underline{d}} \, \eta_{\underline{c} \underline{d}} \\
& + h_{\underline{a} \underline{b}} 
{
+  G \, d_{\underline{a}}^{\,\;\underline{c}}\, h_{\underline{b} \underline{c}}
+  G \, d_{\underline{b}}^{\,\;\underline{c}}\, h_{\underline{a} \underline{c}}
}
+ G^2 \, d_{\underline{a}}^{\,\;\underline{c}}\, d_{\underline{b}}^{\,\;\underline{d}}\, h_{\underline{c} \underline{d}}
.
\end{split}
\end{equation}

Let us use $j_{\underline{a}}^{\,\;\underline{c}}$ to denote minus half the antisymmetric part
of $d_{\underline{a}}^{\,\;\underline{c}}$; namely,
$j_{\underline{a}\underline{b}} \equiv j_{\underline{a}}^{\,\;\underline{c}} \, \eta_{\underline{b} \underline{c}}
=-j_{\underline{b}}^{\,\;\underline{c}} \, \eta_{\underline{a} \underline{c}}$;
the one has
\begin{equation}
  G \, d_{\underline{a}}^{\,\;\underline{c}}\, \eta_{\underline{b} \underline{c}}
=
- \frac{1}{2} h_{\underline{a} \underline{b}} 
- \frac{1}{2} j_{\underline{a}}^{\,\;\underline{c}} \, \eta_{\underline{b} \underline{c}}
+ \mathcal{O}(G^2) ;
\end{equation}
or alternatively
\begin{equation}
 G \, d_{\underline{a}}^{\,\;\underline{c}}
=
-\frac{1}{2} h_{\underline{a} \underline{b}} \, \eta^{\underline{b} \underline{c}} 
-\frac{1}{2} j_{\underline{a}}^{\,\;\underline{c}}
+ \mathcal{O}(G^2) .
\end{equation}
For the first basis vector, one has
\begin{equation}\label{eq:ele-g}
\begin{split}
 l =& \, e_1 
=  \mathbf{e}_{1} 
+ G \, d_{1}^{\,\;\underline{b}} \; \mathbf{e}_{\underline{b}} 
=
  \mathbf{e}_{1} 
- \frac{1}{2} 
(h_{1 \underline{b}} + j_{1 \underline{b}})\, \eta^{\underline{b} \underline{c}}\; \mathbf{e}_{\underline{c}} \\
=&
 \, \mathbf{l} 
- \frac{1}{2} \bigg(
  h_{ll} \, \mathbf{n} 
+ h_{ln} \, \mathbf{l} 
- h_{lm} \, \bar{\mathbf{m}}
- h_{l\bar m} \, \mathbf{m} 
\bigg) \\
&\;\; - \frac{1}{2} \bigg(
  j_{ll} \, \mathbf{n} 
+ j_{ln} \, \mathbf{l} 
- j_{lm} \, \bar{\mathbf{m}}
- j_{l\bar m} \, \mathbf{m} 
\bigg)
.
\end{split}
\end{equation}
And similarly for the other null tetrad vectors one also has:
\begin{equation}\label{eq:ene-g}
\begin{split}
 n =&
 \, \mathbf{n} 
- \frac{1}{2} \bigg(
  h_{nl} \, \mathbf{n} 
+ h_{nn} \, \mathbf{l} 
- h_{nm} \, \bar{\mathbf{m}}
- h_{n\bar m} \, \mathbf{m} 
\bigg) \\
&\;\; - \frac{1}{2} \bigg(
  j_{nl} \, \mathbf{n} 
+ j_{nn} \, \mathbf{l} 
- j_{nm} \, \bar{\mathbf{m}}
- j_{n\bar m} \, \mathbf{m} 
\bigg)
,
\end{split}
\end{equation}
\begin{equation}\label{eq:eme-g}
\begin{split}
 m =&
 \, \mathbf{m} 
- \frac{1}{2} \bigg(
  h_{ml} \, \mathbf{n} 
+ h_{mn} \, \mathbf{l} 
- h_{mm} \, \bar{\mathbf{m}}
- h_{m\bar m} \, \mathbf{m} 
\bigg) \\
& - \frac{1}{2} \bigg(
  j_{ml} \, \mathbf{n} 
+ j_{mn} \, \mathbf{l} 
- j_{mm} \, \bar{\mathbf{m}}
- j_{m\bar m} \, \mathbf{m} 
\bigg)
,
\end{split}
\end{equation}
and
\begin{equation}\label{eq:emeb-g}
\begin{split}
\bar m =&
 \,\bar{ \mathbf{m}}
- \frac{1}{2} \bigg(
  h_{\bar ml} \, \mathbf{n} 
+ h_{\bar mn} \, \mathbf{l} 
- h_{\bar mm} \, \bar{\mathbf{m}}
- h_{\bar m\bar m} \, \mathbf{m} 
\bigg) \\
& - \frac{1}{2} \bigg(
  j_{\bar ml} \, \mathbf{n} 
+ j_{\bar mn} \, \mathbf{l} 
- j_{\bar mm} \, \bar{\mathbf{m}}
- j_{\bar m\bar m} \, \mathbf{m} 
\bigg)
.
\end{split}
\end{equation}

It can be seen that this can easily be extended to second order
by generalizing (\ref{eq:tetrad-relations}) to
\begin{equation}\label{eq:tetrad-relations2}
 e_{\underline{a}} = \mathbf{e}_{\underline{a}} 
+ G \, d_{(1)\underline{a}}^{\;\;\;\;\;\;\underline{b}} \; \mathbf{e}_{\underline{b}} 
+ G^2 \, d_{(2)\underline{a}}^{\;\;\;\;\;\;\underline{b}} \; \mathbf{e}_{\underline{b}} 
+ \mathcal{O}(G^3)
. 
\end{equation}

We could also
define the  the co-basis $f^{\underline{a}}$ and $\mathbf{f}^{\underline{a}}$
from
\begin{equation}
 f^{\underline{a}}_{\; a} \equiv 
\eta^{\underline{a}\underline{b}} \, e_{\underline{b}}^{\;c} \, g_{ca} ,
\end{equation}
and
\begin{equation}
\mathbf{f}^{\underline{a}}_{\;a} \equiv 
\eta^{\underline{a}\underline{b}} \, e_{\underline{b}}^{\;c} \, \eta_{ca} .
\end{equation}

In terms of the co-basis notation one would express the departure from
the flat basis in terms of the quantity $\Delta f^{\underline{a}}_{\,\;\underline{b}}$
given by
\begin{equation}\label{eq:cotetrad-relations}
 f^{\underline{a}} = \mathbf{f}^{\underline{a}} 
+ \Delta f^{\underline{a}}_{\,\;\underline{b}} \; \mathbf{f}^{\underline{b}} .
\end{equation}
Then in first order one has
\begin{equation}
\begin{split}
\delta_{\underline{a}}^{\;\underline{b}} =& e_{\underline{a}} f^{\underline{b}} \\
=&
\big(
\mathbf{e}_{\underline{a}} 
+ G \, d_{(1)\underline{a}}^{\;\;\;\;\;\;\underline{c}} \; \mathbf{e}_{\underline{c}} 
\big)
\big(
\mathbf{f}^{\underline{b}} 
+ \Delta f^{\underline{b}}_{\,\;\underline{d}} \; \mathbf{f}^{\underline{d}} \big) \\
=&
\delta_{\underline{a}}^{\;\underline{b}}
+ G \, d_{(1)\underline{a}}^{\;\;\;\;\;\;\underline{b}}
+ \Delta f^{\underline{b}}_{\,\;\underline{a}} ;
\end{split}
\end{equation}
which means that

\begin{equation}\label{eq:cotetrad-relations}
 f^{\underline{a}} = \mathbf{f}^{\underline{a}} 
- G \, d_{\underline{b}}^{\,\;\underline{a}} \; \mathbf{f}^{\underline{b}} ;
\end{equation}
and also note that $f^{1,2,3,4}=(n,l,-\bar m, - m)$;
where we are omitting the abstract index denoting the one-form
character of the expressions.

We are using the implicit notation in which, with the exception of the expressions
$e_{\underline{a}}$, the indices $\underline{a}$ refer to components with respect
to the tetrad vectors $\mathbf{e}_{\underline{a}}$.

Then, one could also work with this co-basis for the calculations;
but we will continue with the vectorial representation.

\subsection{Relation with the asymptotic structure}

In the study of asymptotically flat spacetimes one normally recurs to the
conformal techniques so that the conformal metric
\begin{equation}
 \tilde g_{ab} =  \Omega^2 g_{ab} ,
\end{equation}
is regular at future null infinity.
One can then define the regular tetrad at infinity given by\cite{Moreschi86}
\begin{align}
 \hat l^a &= \Omega^{-2} l^a \label{eq:hatl} \\
 \hat m^a &= \Omega^{-1} m^a \label{eq:hatm} \\
 \hat n^a &=  n^a \label{eq:hatn} .
\end{align}

The asymptotic structure provides several ways in which one can decompose the 
asymptotic metric in terms of a flat background $\eta_{\text{asy}}$
and a deviation term $h_{\text{asy}}$; so that
\begin{equation}\tag{\ref{eq:asymp1}}
 g = \eta_{\text{asy}} + h_{\text{asy}} .
\end{equation}

Let us assume for a moment that we have one such decomposition.

For those decompositions, one can also define
\begin{equation}
 \tilde \eta_{ab} =  \Omega^2 \eta_{ab} ;
\end{equation}
which must coincide asymptotically with $\tilde g_{ab}$.
In other words, the components of $\Omega^2 h_{ab}$ in terms of any regular tetrad
must go to zero like $\Omega$.
Equivalently we may express
\begin{equation}
\Omega^2  h_{\hat{\underline{a}}\hat{\underline{b}}}
=
\frac{1}{r}  h_{(0)\hat{\underline{a}}\hat{\underline{b}}}
+
\frac{1}{r^2}  h_{(1)\hat{\underline{a}}\hat{\underline{b}}}
+
\frac{1}{r^3}  h_{(2)\hat{\underline{a}}\hat{\underline{b}}}
+ \ldots
\end{equation}
where the indices $\hat{\underline{a}},\hat{\underline{b}},...$ refer to a regular
tetrad components and $r$ is an appropriate asymptotic radial coordinate.

Some specific tetrad components are:
\begin{equation}
\Omega^2  h_{\hat{l}\hat{l}}
=
\Omega^{-2}  h_{l l}
=
\frac{1}{r}  h_{(0)\hat{l }\hat{l }}
+ \ldots
=
\Omega^{-4}
\frac{1}{r}  h_{(0)l l}
+ \ldots
,
\end{equation}

\begin{equation}\label{eq:heleeme}
\Omega^2  h_{\hat{l}\hat{m}}
=
\Omega^{-1}  h_{l m}
=
\frac{1}{r}  h_{(0)\hat{l }\hat{m }}
+ \ldots
=
\Omega^{-3}
\frac{1}{r}  h_{(0)l m}
+ \ldots
,
\end{equation}

\begin{equation}\label{eq:hemeeme}
\Omega^2  h_{\hat{m}\hat{m}}
=
 h_{m m}
=
\frac{1}{r}  h_{(0)\hat{m}\hat{m}}
+ \ldots
=
\Omega^{-2}
\frac{1}{r}  h_{(0)m m}
+ \ldots
,
\end{equation}

\begin{equation}
\Omega^2  h_{\hat{n}\hat{n}}
=
\Omega^{2}  h_{n n}
=
\frac{1}{r}  h_{(0)\hat{n }\hat{n }}
+ \ldots
=
\frac{1}{r}  h_{(0)n n}
+ \ldots
;
\end{equation}
from which we can see that the most asymptotically divergent tetrad component is $h_{n n}$
which could grow as $\Omega^{-1}$.
This means that we could also express
\begin{equation}
\Omega^{2}  h_{{\underline{a}}{\underline{b}}}
=
\frac{1}{r}  h_{[0]{\underline{a}}{\underline{b}}}
+
\frac{1}{r^2}  h_{[1]{\underline{a}}{\underline{b}}}
+
\frac{1}{r^3}  h_{[2]{\underline{a}}{\underline{b}}}
+ \ldots
\end{equation}
where the indices ${\underline{a}},{\underline{b}},...$ refer to a standard tetrad components.
The relations (\ref{eq:hatl})-(\ref{eq:hatn}) provide with the transformations between
the components $h_{(i)\hat{\underline{a}}\hat{\underline{b}}}$ and 
the components $h_{[j]{\underline{c}}{\underline{d}}}$.

When considering the case of Einstein equations of general relativity,
the natural question arises: Do the solutions to the linear problem in standard
gauges have this natural asymptotic decomposition?
One can check by direct calculation, of the Schwarzschild solution 
in the harmonic gauge that, for example, the component
\begin{equation}
 h_{ll} = - \frac{4 M_A}{r} + \mathcal{O}(\frac{1}{r^2}) ;
\end{equation}
which is even true for the Schwarzschild solution in the standard 
coordinate system; where in both cases we take the vector $l$ to be
$l = \frac{\partial }{\partial t} + \frac{\partial }{\partial r}$,
in terms of their intrinsic spherical coordinates.
Therefore we conclude that the answer is negative; and so we do not expect
all of the above relations, for the asymptotic behaviour of the null tetrad components
of the solution $h$, to apply.

However, the Schwarzschild solution is asymptotically flat; so one could
still use the assignment for the first order null tetrad, based
on the given null tetrad of the flat background metric associated to
the solution $h$, as given previously, to calculate all of the 
asymptotic quantities, including radiation field and curvature components.

\subsection[Relation between the interior flat metric and the 
asymptotic flat metrics]
{Relation between the interior flat metric and the 
asymptotic flat metrics}

It is also important to remark that in having a decomposition in the interior of the form
\begin{equation}\tag{\ref{eq:getamash}}
g_{ab} = \eta_{ab} + h_{ab}
;
\end{equation}
in which the flat metric $\eta_{ab}$ does not belong to the family
of adapted asymptotic flat metrics $\eta_{\text{asy}}$;
when one considers two points in the spacetime $(\mathscr{M}_\eta,\eta_{ab})$,
the two future directed null cones define two sections at future
null infinity that it will be related by a BMS translation at
$\mathscr{I}^+_\eta$, of the spacetime $(\mathscr{M},\eta_{ab})$;
but it will not coincide in general with a BMS translation at
future null infinity $\mathscr{I}^+_g$, of the original
spacetime $(\mathscr{M},g_{ab})$.
Instead, these two sections will be related by a supertranslation.

It is because of this reason that an inertial (Bondi) system with respect to the spacetime
structure $(\mathscr{M}_\eta,\eta_{ab})$ might not coincide in general with
inertial systems of the spacetime structure $(\mathscr{M},g_{ab})$.
More concretely, sections $\tilde u=$constant of a inertial system of $(\mathscr{M},\eta_{ab})$,
at the boundary manifold $\mathscr{I}^+$, will be defined by a family 
$\alpha(u_B,\zeta,\bar\zeta)$, in terms of a inertial system of the
asymptotic structure of the spacetime $(\mathscr{M},g_{ab})$.

In the construction of a model for compact objects in  
metric gravitational theories
one is normally forced to relate a choice of an interior flat metric $\eta$
with an asymptotic flat  metric $\eta_{\text{asy}}$; so the question is:
how does one choose the interior flat metric and the asymptotic
flat metric? since the choice seems to be crucial for the construction
of the model.
The answer is related to the general objective of describing the dynamical
evolution of the physical quantities of the system.
In the interior of the spacetime, we have review in section \ref{sec:special-relativity}
the basic concepts of total momentum and total angular momentum;
which in turn comes along with the concept of center of mass.
At future null infinity we have constructed in references \cite{Moreschi04}
and \cite{Gallo:2014jda} the two definitions of intrinsic angular momentum
in general relativity which do not suffer from supertranslation ambiguities.
These definitions of intrinsic angular momentum come along with the 
corresponding definition of center of mass frame at future null infinity.
Using these center of mass frames, we can select, at each observing asymptotic
point $p$ a unique inertial asymptotic flat metric $\eta_{\text{asy}}$.

Since the model for compact objects will intend to relate the value of
physical quantities in the interior with value of physical quantities
in the asymptotic region, one should use in both regimes the frames
which are best related to the \emph{center of mass frames}.
For this reason we next define an asymptotic flat 
metric coming from the definition of center of mass at future null infinity.

\subsection{The asymptotic center of mass flat metric}

We can use the definition of center of mass at future null infinity
to define a new preferred flat asymptotic 
background using the 
invariance of the translations under BMS\cite{Sachs62} transformations.
Let us see this in detail.
The set of center of mass sections\cite{Moreschi04,Gallo:2014jda} at future null
infinity, can be used to define a retarded null coordinate 
$u_{\mathtt{cm}}$. Let denote with $S_{\mathtt{cm}}$, 
the non-intersecting\cite{Moreschi98}, 
one parameter family of center of mass sections as described
in \cite{Moreschi04} or \cite{Gallo:2014jda},
then, the null congruence of null geodesics reaching $S_{\mathtt{cm}}$
orthogonally form a null hypersurface. We define $u_{\mathtt{cm}}$
to be the null function that is constant on these null hypersurfaces
and which increment adjusts to the time generator 
$t^a = \frac{\mathcal{P}^a}{M}$, where $\mathcal{P}^a$ is the total momentum
and $M$ the total mass.
We can refer this retarded null coordinate to the retarded null coordinate $\tilde u$
of an asymptotic inertial frame with:
\begin{equation}\label{eq:dutildeducm}
 \frac{\partial \tilde u}{\partial u_{\mathtt{cm}}} 
= V_{\mathtt{cm}}(u_{\mathtt{cm}},\zeta,\bar\zeta) > 0 ,
\end{equation}
where the time derivative of the inertial (Bondi) time $\tilde u$ 
with respect to the non-inertial center of mass retarded null time $u_{\mathtt{cm}}$
is expressed in terms of the scalar $V_{\mathtt{cm}}(u_{\mathtt{cm}},\zeta,\bar\zeta)$.

If one makes a decomposition of the scalar $V_{\mathtt{cm}}(u_{\mathtt{cm}},\zeta,\bar\zeta)$
in terms of spherical harmonics; the first four terms, with $l=0,1$,
have the information of an infinitesimal translation, when the coordinate
$u_{\mathtt{cm}}$ is incremented. 
But this information, the infinitesimal translations,
is invariant under actions of the BMS group up to order $\lambda$.
This means that we can use this information to define a
\emph{center of mass asymptotic flat background}.
Let us define $V_I$ as the projection of the scalar $V_{\mathtt{cm}}(u_{\mathtt{cm}},\zeta,\bar\zeta)$
to the subspace of functions with spherical harmonic expansion only involving
the $l=0,1$ terms.
Then, the flat metric $\eta_{\mathtt{cm}}$ defined from the line element:
\begin{equation}
ds^2 =
 (1 - 2 r \frac{\dot V_I}{V_I} )\, du_{\mathtt{cm}}^2 + 2\, du_{\mathtt{cm}} \, dr_{\mathtt{cm}} 
- \frac{r_{\mathtt{cm}}^2}{V_I^2} dS_0^2 
;
\end{equation}
where $dS_0^2 = d\zeta \, d\bar\zeta / P_0^2$ is the metric of the unit sphere,
and 
$\dot V_I = \frac{\partial V_I }{\partial u_{\mathtt{cm}} }$;
is invariant up to order $\lambda$ under BMS actions.
The natural definition of $r_{\mathtt{cm}}$ comes from the requirement to 
agree asymptotically with the luminosity distance.

Notice that this flat asymptotic metric $\eta_{\mathtt{cm}}$ 
can only coincide with
an inertial (Bondi) asymptotic flat metric, in the case of absence of
gravitational radiation.
Its definition is dependent on the dynamical evolution of the spacetime.

Returning to the relation of frames in the interior with those of the asymptotic
region; the use of the interior flat metric $\eta$ depends on the particular
gauge set up that is chosen, but normally all the physically useful
variables are referred to the center of mass frame\cite{Moreschi95} available.
Due to the complicated structure of the BMS asymptotic symmetry group,
involving the infinite dimensional supertranslation freedom,
one must associate the interior defined physical variables,
with those defined in the asymptotic region which are free from
supertranslation gauge problems.
For this reason one should use physical variables defined with respect
to the center of mass frame defined in the asymptotic 
region\cite{Moreschi04,Gallo:2014jda}, and should also relate
the interior flat metric $\eta$ with the unique  asymptotic flat 
metric $\eta_{\mathtt{cm}}$, presented above.
Failing to make this relation of physical variables, 
will involve unnoticed dependence on supertranslation
gauge issues, with unclear physical power of predictability and
description\cite{Moreschi88b}.

\subsection{The radiation fields in a linear spacetime}

Here we consider the metric $g_A$, which is assumed to be asymptotically flat,
and therefore it can be decomposed in the form (\ref{eq:asymp1}).
We will use the set of null tetrads described in subsection \ref{subsec:general}
and study the radiation fields for large values of affine distance along future null directions
of the flat background metric.

Our calculation of the shear gives:
\begin{equation}
 \sigma_0 = -\frac{1}{2} \big( \eth_0(h_{0lm}) + \eth_0(j_{0lm}) +  h_{0mm} \big) .
\end{equation}
It must be emphasized that our result has an extra term 
involving the edth operator when compared with equation 46  of
reference \cite{Walker:1979zk}.
These computations have been checked with {\tt REDUCE}.

We have indicated before that in these kind of approximations one should not focus on the 
concept of spin coefficient $\sigma$, but instead should consider as more
representative of the radiation content the spin coefficient $\sigma'$.
In particular, its asymptotic leading order behavior $\sigma'_0$
is independent of the gauge $j_{\underline{a}\underline{b}}$
 discussed above, and one always has
\begin{equation}\label{eq:sigma'0a}
 \sigma'_0 = \frac{1}{2} \frac{\partial h_{0\bar m \bar m}}{\partial \tilde u} ;
\end{equation}
and
\begin{equation}
 \Psi_4^0 = \frac{\partial \sigma'_0}{\partial \tilde u} ;
\end{equation}
where $\tilde u$ is an asymptotic inertial null retarded coordinate.

These then are the two fields that codify the appearance of gravitational radiation.

\section{The appropriate dynamical times and 
the balanced \\
equations of motion}\label{sec:times}

In this section we gather together the main ideas and use them to build the 
general form of the balanced equations of motion.

Let us recall the metrics we have been using.
We assume that there exists an exact metric of the spacetime of the form
given in (\ref{eq:gAB}).
We study in detail the asymptotic structure of the sub-metric $g_A$ (\ref{eq:gA})
by disregarding the effects of the complementary system $B$.
We adjust $g_A$ to model a compact object, which dynamics
is affected by the existence of system $B$, and therefore
by the geometry determined by the sub-metric $g_B$, appearing in (\ref{eq:gB}).

In the calculation of the back reaction effects one models
the emited gravitational radiation by considering the asymptotic
radiation fields in the metric $g_A$; although of course in the real
world, the radiation emitted by body A will be partially absorbed
by system B, and therefore will never reach future null infinity.
But in any case we do choose to estimate the local back reaction
effects by this method; since it is difficult to achieve a
local definition
of gravitational radiation emitted by a compact object.

\subsection{Form of the dynamical equations in the interior}

The basic dynamics is based on the existence of a smooth curve
$C(t)$ in the manifold, such that it is 
timelike with respect to the metrics  $\eta$ and $g_B$.

In analogous studies of the dynamics of charged particles in Minkowski spacetime,
we have found\cite{Gallo:2011tf} that new degrees of freedom might be expected.
In that case, the dynamics could be presented in an equations of motion of the form
\begin{equation}
 \mathcal{D}(v,\dot v,\ddot v, \alpha_1, \alpha_2) = \mathcal{F}(x^\mu,{v}, \dot v) ;
\end{equation}
where the $\alpha$'s where new degrees of freedom, and with
\begin{equation}
 g(v,\mathcal{F}) \neq 0 .
\end{equation}

In the gravitational case, the balanced equations of motion is constructed by correcting, 
in first order
of gravitational radiation, the dynamics dictated by (\ref{eq:dpdtau});
where the new right hand side will have the information of the total 
radiated momentum. 
But observing the balanced equation (\ref{eq:bondibalanceV}) at future null infinity
one can notice that the timelike component is not vanishing.

Therefore, 
in the interior, the balanced equations of motion must have (at least) the structure
\begin{equation}\label{eq:balanced}
M_A \Big( v^a \nabla_{(B)a} \, v^b  + w v^b \Big)(u_\text{int}) 
= {F}^b(u_\text{asy}) ;
\end{equation}
where $w$ shows a new possible degree of freedom,
and it is emphasized that the left hand side is expressed
in terms of an interior notion of time $u_\text{int}$,
that we will fix below, and 
the right hand side is expressed in terms of a
corresponding asymptotic notion of dynamical 
time $u_\text{asy}$,
and where $F$ is minus the integrated flux of gravitational radiation
on an asymptotic sphere; namely
\begin{equation}\label{eq:force}
 F^\mu 
 = -\mathcal{F}_V^{\mu}
 = - \frac{1}{4 \pi } \int_S 
{\hat l_0}^{\mu} \tilde V \, \sigma'_0 \, {\bar \sigma}'_0 \, dS^2 
.
\end{equation}
The final relation between $u_\text{asy}$ and $u_\text{int}$ will be
settled in subsection \ref{subsec:interiorandexterior} below.

Although the natural dynamical time for the seed equation (\ref{eq:dpdtau})
is $\tau$, for the actual numeric computation 
we can also use as parameterization the proper time $\tau_0$,
with respect to the metric $\eta$.
Let $\mathbf{v}$ and $v$ be the corresponding tangent vectors
to the proper times $\tau_0$ and $\tau$.
Then the basic differential operators are $\mathbf{v}^b \partial_b \mathbf{v}^a$
or ${v}^b \nabla_{(B)b} {v}^a$;
where $\partial_b$ is the covariant derivative associated with the metric $\eta$.

Let us note that the two velocity vectors are proportional
\begin{equation}
 v^b = \Upsilon \mathbf{v}^b .
\end{equation}
So that one has
\begin{equation}
 1 = g_B(v,v) = \Upsilon^2 g_B(\mathbf{v},\mathbf{v})
= \Upsilon^2 \big( 1 +  h_B(\mathbf{v},\mathbf{v}) \big)
 ;
\end{equation}
which gives $\Upsilon$ in terms of $\mathbf{v}$ and $g_B$.

Expressing the covariant derivative $\nabla_{(B)a}$ of $g_B$, in terms of the covariant 
derivative $\partial_a$ of $\eta$, one has
\begin{equation}
 \nabla_{(B)a} \, v^b = \partial_a \, v^b + \gamma^{\;b}_{a\;\;c} \,  v^c 
;
\end{equation}
and using the relation between the vectors $v$ and $\mathbf{v}$ one also has
\begin{equation}
\begin{split}
v^a \nabla_{(B)a} \, v^b = 
\Upsilon^2 \mathbf{v}^a \partial_a \, \mathbf{v}^b 
 + \Upsilon \frac{d\Upsilon}{d\tau_0} \, \mathbf{v}^b
 + \Upsilon^2 \gamma^{\;b}_{a\;\;c} \, \mathbf{v}^a \,  \mathbf{v}^c 
.
\end{split}
\end{equation}

Let us introduce the notation:
\begin{equation}
\mathbf{a}^a \equiv \mathbf{v}^b \partial_b \mathbf{v}^a ,
\end{equation}
and
\begin{equation}
\begin{split}
\mathbf{f}^b 
\equiv
-  \gamma^{\;b}_{a\;\;c} \, \mathbf{v}^a \,  \mathbf{v}^c 
- \frac{1}{\Upsilon}  \frac{d\Upsilon}{d\tau_0} \, \mathbf{v}^b
- \frac{w}{\Upsilon}  \mathbf{v}^b
;
\end{split}
\end{equation}
so that the equations of motion can be expressed as
\begin{equation}\label{eq:mot-compact}
\mathbf{a}^a =  \mathbf{f}^a + f_\lambda^a .
\end{equation}
with $ f_\lambda^\mu$ defined by
\begin{equation}\label{eq:flambda-F0}
M f_\lambda^\mu \equiv \frac{1}{\Upsilon^2} F^\mu
.
\end{equation}	

We also define $\mathbf{f}_\perp^b$ by
\begin{equation}
\mathbf{f}_\perp^b \equiv - \gamma^{\;d}_{a\;\;c} \, \mathbf{v}^a \,  \mathbf{v}^c 
\big(
\eta_d^{\;\;b} - \mathbf{v}_d \mathbf{v}^b
\big)
;
\end{equation}
which, it should be remarked, only depends on the background 
geometry $g_B$ and $\mathbf{v}$.

\subsection{Relation between the interior and exterior dynamics}\label{subsec:interiorandexterior}

In general, the charge integrals of the Riemann tensor give a connection
between the asymptotically defined total momentum $\mathcal{P}$
and the momentum $P$ defined in the interior for the particle.
This relation is given in terms of a metric structure just determined by $g_A$.
But the dynamics we are looking for involves the interaction
of particle $A$ with the background metric $g_B$; which generates
gravitational radiation that affects the dynamics.

In particular, the components of the momentum are calculated with respect
to the generators of translations; which at infinity are the BMS translations,
and in the interior are translations of the background flat metric $\eta$.
It is a fact that when effects of gravitational radiation are taken into 
account, the interior flat metric $\eta$ does not coincide with any
of the infinite family of flat metrics $\eta_\text{asy}$, which give 
the inertial frames at future null infinity.
Therefore, the way in which we identify translations at infinity
with translations in the interior, will not only depend on gauge
choices but also necessarily introduce
effects from gravitational radiation. Since our equations of motion
is quadratic on these effects, in the lower order approach
we will discard terms of cubic and higher order involving gravitational
radiation.

The way in which the exact expression for the left hand side of (\ref{eq:balanced}) is related to the 
left hand side of (\ref{eq:bondibalanceV}) depends on the detail of the model
one is making for the interior of the  spacetime.
In any case, independently from the gauge choice one is making in the interior,
we have shown before that in order to avoid supertranslation ambiguities
one should relate physical quantities in the asymptotic region with respect
to the asymptotic center of mass frame,
with the physical quantities in the interior region with respect
to the interior center of mass frame.

As we have just noted, the notion of total momentum
described at future null infinity, can be related to the interior notion
of momentum through the linear structure of the metric $g_A$ with 
respect to the flat background metric $\eta$, in the interior
of the spacetime. 
But the time derivative of this quantity might depend on the
detailed dynamics in the interior, with higher order information.

Let us note that $F^\mu$ is expressed in terms of 
an asymptotic retarded time and therefore as a function  of $\tilde V$; 
which, in terms of an interior dynamical time, it can be written as
\begin{equation}\label{eq:dutildedtau}
\tilde V(u,\zeta,\bar\zeta)
=
\frac{\partial \tilde u}{\partial u} 
= \frac{1}{ \big(\frac{du}{d\tau}\big) }
\frac{\partial \tilde u}{\partial \tau} 
= \frac{1}{ \big(\frac{du}{d\tau}\big) }
V(\tau,\zeta,\bar\zeta) ;
\end{equation}
with the new notation
\begin{equation}\label{eq:ve}
V =  \frac{\partial \tilde u}{\partial \tau} .
\end{equation}

Then, we can define the gravitational radiation force,
in terms of the $\tau$ time, by 
\begin{equation}\label{eq:force_tau}
\mathbf{F}^\mu(\tau)
= - \frac{1}{4 \pi } \int_S 
{\hat l}_0^{\mu} \ V \, \sigma'_0 \, {\bar \sigma}'_0 \, dS^2 
.
\end{equation}
This also invites to define the gravitational radiation force,
with respect to the proper time $\tau_0$;
namely:
\begin{equation}\label{eq:force_tau_0}
\mathbf{F}_0^\mu(\tau_0)
= - \frac{1}{4 \pi } \int_S 
{\hat l}_0^{\mu} \ V_0 \, \sigma'_0 \, {\bar \sigma}'_0 \, dS^2 
;
\end{equation}
where we use the definition
\begin{equation}\label{eq:ve_0}
V_0 =  \frac{\partial \tilde u}{\partial \tau_0} .
\end{equation}
Note that $\mathbf{F}^\mu = \Upsilon \mathbf{F}_0^\mu$.

After considering the general form of the balanced
equations of motion in the form
\begin{equation}\tag{\ref{eq:balanced}}
M_A \Big( v^a \nabla_{(B)a} \, v^b  + w v^b \Big)(u_\text{int}) 
= {F}^b(u_\text{asy}) ;
\end{equation}
we choose to express the right hand
side in terms of the interior dynamical time.
With this choice, the left hand side is already expressed in terms of
its natural dynamical time; which simplifies the algebraic expressions,
and one needs only to consider the details of expressing
the flux force in terms of the interior time.
As explained above one should allow for a degree of freedom
that relates the asymptotic time with the interior one;
which it can be encoded in the quantity $\chi=\frac{du}{d\tau}$.
Then we express
\begin{equation}\label{eq:balanced-int}
M_A \Big( v^a \nabla_{(B)a} \, v^b  + w v^b \Big)(\tau) 
= {F}^b(u_\text{asy})
= \frac{1}{\chi} \mathbf{F}^\mu 
= \frac{\Upsilon}{\chi}  \mathbf{F}_0^\mu
.
\end{equation}

Using the $\tau_0$ dynamical time on the let hand side,
one can express the equations of motion
by
\begin{equation}\label{eq:balanced-int_0}
M_A \bigg(
 \mathbf{v}^a \partial_a \, \mathbf{v}^b 
+  \gamma^{\;b}_{a\;\;c} \, \mathbf{v}^a \,  \mathbf{v}^c
+ 
\Big(
\frac{1}{\Upsilon}  \frac{d\Upsilon}{d\tau_0}
+
\frac{w}{\Upsilon}
\Big)
 \, \mathbf{v}^b
\bigg)
=
\frac{1}{\chi\Upsilon}  \mathbf{F}_0^\mu
.
\end{equation}

It is important to remark that this natural consideration for a
degree of freedom relating interior and asymptotic dynamical
times, is not necessary in the null gauge; as we will
show in a separate work.

Thus, equation (\ref{eq:balanced-int_0}) is the main equations of motion,
as expressed with respect to $\tau_0$.
Contracting this equation with $\eta_{bd} \mathbf{v}^d$ gives
\begin{equation}\label{eq:balanced-int_0-con-v}
\begin{split}
M_A \Big(&
\gamma^{\;b}_{a\;\;c} \, \mathbf{v}^a 
\,  \mathbf{v}^c \, \eta_{bd} \mathbf{v}^d
+
\frac{1}{\Upsilon}  \frac{d\Upsilon}{d\tau_0}
+
\frac{w}{\Upsilon}
\Big) 
=
\frac{1}{\chi\Upsilon}
\, \mathbf{F}_0^b \eta_{bd} \mathbf{v}^d
;
\end{split}
\end{equation}
and it remains the equations of motion
\begin{equation}\label{eq:balanced-int_0-flat}
\begin{split}
M_A \,
\mathbf{a}^b
= M_A \, \mathbf{f}_\perp^b
+ 
\frac{1}{\chi\Upsilon}
\, \mathbf{F}_0^d \;
\big(
\eta_d^{\;\;b} - \mathbf{v}_d \mathbf{v}^b
\big)
.
\end{split}
\end{equation}

Equation (\ref{eq:balanced-int_0-con-v}) is understood as an equation for $w$.
The main dynamical equation is then (\ref{eq:balanced-int_0-flat});
where the possible degree of freedom $\chi$ depends on the detail
nature of gauge being used. As we have said, in the case
of the null gauge that we will present in a separate article, one
has that $\chi=1$.

\subsection{The balanced equations of motion in terms of orders}

Our goal is to set the equations of motion for a compact object
(a particle) that takes into account the first order back reaction
due to the emission of gravitational radiation.
This requires to deal with the dynamics in the interior of the spacetime
and also with the gravitational radiation defined at future null infinity.

In setting the interior geometry we have found a natural expansion in terms
of the gravitational constant $G$. 

In the asymptotic study of gravitational radiation, it appears very
naturally an expansion in terms of a quantity $\lambda$ 
that has the information on the strength of gravitational radiation,
and is defined such that the flux of momentum is of order $\lambda^2$.

In making a link between the interior study and the asymptotic study, it is then
natural to make a double expansion in terms of powers of $G$ and $\lambda$.
We will therefore refer to a $(p,q)$ order meaning terms up to order
$(G^p,\lambda^q)$.

Someone could say that it is unnecessary to include in the discussion of
orders the $\lambda$ parameter; since actually
one could estimate the way in which the gravitational constant appears
in the radiation field.
In particular, one can deduce that since at lowest order the radiation field
is calculated from the term $h_A$ of the metric, it is already of
order $G$. But also, since the radiation field is generated by 
the accelerated motion of the particle in the background, it is also
of order $M_B$; which means  
that $\lambda = \mathcal{O}(M_A M_B) = \mathcal{O}(G^2)$.
Although this is true, we prefer to keep the control parameter
$\lambda$ for the discussion of orders; since it better captures 
the incidence of the radiation fields in the dynamics,
and also, because from a formal point of view, the accelerations
could be from non-gravitational origin, as electromagnetic one,
and we should be able to discuss the motion for those cases too.
So we only assume that $\lambda = \mathcal{O}(G)$.

The functional form of {$\mathbf{F}_0$} depends crucially on the gauge
one is using to relate the interior dynamics with the asymptotic region.
So, it might be that actually 
$\mathbf{F}_0 = \mathbf{F}_0(u_\text{asy}, \mathbf{v} , \dot{\mathbf{v}})$;
that is, the flux might also depend on time derivatives of the velocity.
Because of this we next indicate the way the equations of motion must
be understood.

Let us observe that at first sight, the first term on the left hand side
of (\ref{eq:balanced-int_0})
is of order (1,0): since $M_A$ is order $G$.
The other terms
on the right hand side could be considered of order (1,0) 
in the non-gravitational case, or of order (2,0) in the binary case.
The right hand side depends quadratically on the radiation and
therefore is at least of order (2,2).

The idea is then, to consider first the equation up to order $G^2$, and calculate
the acceleration at this order; and then we proceed to
calculate the flux of gravitational radiation, that will depend on the 
details of the chosen gauges, and with this, we build the complete
equation and integrate the operator $\mathbf{v}^a \partial_a \, \mathbf{v}^b$
at order (2,2).
To work at orders of the exterior forces meas that,
any possible appearance of a time derivative on the velocity,
in the flux, must be replaced by  $\mathbf{f}_\perp$; so that
in the general case, $\mathbf{F}_0$ must be understood as 
\begin{equation}
 \mathbf{F}_0 = \mathbf{F}_0({\tau_0}, \mathbf{v},  \mathbf{f}_\perp) .
\end{equation}

As indicated before, the details of these expressions, depend on
the preferred gauge choices.

Of course, for a specify use of the equations of motion one should
maintain the different terms at consistent orders of approximation.

\section{Final comments}\label{sec:comments}

When embarking on the construction of a model for compact objects within the particle
paradigm in general relativistic theories, it is natural to recur to approximation schemes in
terms of some order parameter. 
One knows from the beginning that such construction will have
sense only if it is thought in terms of finite orders.

In this work we have presented the general form of the equations of motion for 
particles subjected to back reaction effects due to gravitational radiation by
using various choices of dynamical time.
We have also pointed out several of the issues one must handle when
constructing balanced equations of motion, as for example the relation
of generators of translations in the interior an at future null infinity.
These type of problems is generally overlooked in the literature.
The reasons for the usual simplifications of the descriptions is 
because the subject is very complex, as has been indicated in \cite{Moreschi88b}.
The specific equations of motion can be calculated when a definite
field equation and a choice of gauges has been made.

Our main basic working assumptions have been:
that the back reaction effects due to gravitational radiation can be well represented
by the calculation in the 
leading order effects
of the particle
using the unaffected null cones;
that the notion of center of mass at future null infinity are well described by
the previous works of the authors;
that to each point  at future null infinity there corresponds a 
center of mass
flat
background metric, as described above, 
which must be related to the
corresponding notion of center of mass in the interior at the needed order
of approximation;
and that the first correction to the zero order equations of motion
comes from the balance of the change of momentum with the radiation
of momentum at infinity.
In relation to the structure of the particle, we have assumed that
its mechanical properties are completely determined by a constant mass
and we have disregarded possible spin and dipole contributions.
Regarding the dipole, we have remarked in section \ref{subsec:intrinsic}
that at zero order, particles should not be assigned any dipole structure.

With respect to the unspecified theory of gravity that describes
the structure of the spacetime,
we recall here that in this presentation we are assuming:
that isolated objects are well represented by asymptotically flat spacetimes;
that the linear structure of the field equation is such that allows for the
identification of the total momentum at infinity with the total momentum
calculated in the interior from the energy momentum tensor;
that the energy momentum tensor is conserved, that is its divergence is zero.

Some aspects of the model presented here has certain similarities with the so 
called post-Minkowskian formalisms usually studied within the framework of Einstein's 
equations\cite{Bel:1981, Havas:1965, Rosemblum:1983}. However, our approach differs 
from those formalisms in the way in which the equations of motion are obtained because our 
approach makes use of the energy-momentum tensor and its conservation rather than 
obtaining them through particular field equations. Furthermore, the post-Minkowskian 
formalism is usually presented in the harmonic gauge, while our presentation is 
independent of the chosen gauge, even thought the fine details of the equations 
of motion depend on each choice of gauge and field equations under consideration.

To all this, we can add that the way in which different gauges affect the final 
form of the equations of motion, is a subject that deserves a detailed study
that we will present elsewhere. 
In particular, it should be stressed that our result has been calculated
with the leading order geometry for the object and background; since we
have been concerned with the first order back reaction, due to gravitational
radiation. In order to improve in the predicting power of the equations,
then one should extend the calculations to higher orders of
the specific field equations, that are being used.
For example, in \cite{Gallo:2018a} we present the balanced equations
of motion in general relativity in the null gauge, up to second order
in the accelerations; 
and in \cite{Gallo:2017yys} we present the
corresponding balanced equations of motion in the harmonic gauge
in first order of the accelerations and field equations.

It is important to remark that in analogy to what we have found in
the case of charged particles in Minkowski spacetime, there are 
at least two further
degrees of freedom involved in the dynamics.
Although the $w$ degree of freedom might be associated to whether one wants to
consider a constant mass parameter or not for the particle; the 
best choice for $w$ depends on the gauge one has chosen.

Furthermore, particular gauge choices might involve new degrees
of freedom; as we will show in separate works.

The fact that the final form of the equations of motion is dependent on the 
choice of gauges for the calculations, should not be understood in a
negative way; on the contrary, it is the natural expectation for a
finite degree of freedom model intended to represent 
and infinite degree of freedom system.
However from the astrophysical point of view there is a general believe
that the main observational properties in the collapsing phase of compact
object system, might be well represented by models with finite degrees
of freedom.
From another point of view, models with finite degrees of freedom might
provide well approximated initial data to be used in full numerical
calculations for the final stages of collapse.

This approach to the equations of motion can be generalized in a number of ways. 
The natural next topic
that we plan to deal is the introduction of spin for the particles. 
This and other issues will be dealt in future works.

\subsection*{Acknowledgments}

We have benefited from discussions with R. Wald, to whom we are
very grateful.
We acknowledge support from CONICET and SeCyT-UNC. 

%
%
%

\begin{thebibliography}{10}
	
	\bibitem{Abbott:2016blz}
	{\bfseries Virgo, LIGO Scientific} Collaboration, B.~P. Abbott {\em et~al.},
	``{Observation of Gravitational Waves from a Binary Black Hole Merger},''
	\href{http://dx.doi.org/10.1103/PhysRevLett.116.061102}{{\em Phys. Rev.
			Lett.} {\bfseries 116} no.~6, (2016) 061102},
	\href{http://arxiv.org/abs/1602.03837}{{\ttfamily arXiv:1602.03837 [gr-qc]}}.
	
	\bibitem{TheLIGOScientific:2016wfe}
	{\bfseries Virgo, LIGO Scientific} Collaboration, B.~P. Abbott {\em et~al.},
	``{Properties of the Binary Black Hole Merger GW150914},''
	\href{http://dx.doi.org/10.1103/PhysRevLett.116.241102}{{\em Phys. Rev.
			Lett.} {\bfseries 116} no.~24, (2016) 241102},
	\href{http://arxiv.org/abs/1602.03840}{{\ttfamily arXiv:1602.03840 [gr-qc]}}.
	
	\bibitem{Abbott:2016nmj}
	{\bfseries Virgo, LIGO Scientific} Collaboration, B.~P. Abbott {\em et~al.},
	``{GW151226: Observation of Gravitational Waves from a 22-Solar-Mass Binary
		Black Hole Coalescence},''
	\href{http://dx.doi.org/10.1103/PhysRevLett.116.241103}{{\em Phys. Rev.
			Lett.} {\bfseries 116} no.~24, (2016) 241103},
	\href{http://arxiv.org/abs/1606.04855}{{\ttfamily arXiv:1606.04855 [gr-qc]}}.
	
	\bibitem{Abbott:2017vtc}
	{\bfseries VIRGO, LIGO Scientific} Collaboration, B.~P. Abbott {\em et~al.},
	``{GW170104: Observation of a 50-Solar-Mass Binary Black Hole Coalescence at
		Redshift 0.2},'' \href{http://dx.doi.org/10.1103/PhysRevLett.118.221101}{{\em
			Phys. Rev. Lett.} {\bfseries 118} no.~22, (2017) 221101},
	\href{http://arxiv.org/abs/1706.01812}{{\ttfamily arXiv:1706.01812 [gr-qc]}}.
	
	\bibitem{Abbott:2017oio}
	{\bfseries Virgo, LIGO Scientific} Collaboration, B.~P. Abbott {\em et~al.},
	``{GW170814: A Three-Detector Observation of Gravitational Waves from a
		Binary Black Hole Coalescence},''
	\href{http://dx.doi.org/10.1103/PhysRevLett.119.141101}{{\em Phys. Rev.
			Lett.} {\bfseries 119} no.~14, (2017) 141101},
	\href{http://arxiv.org/abs/1709.09660}{{\ttfamily arXiv:1709.09660 [gr-qc]}}.
	
	\bibitem{TheLIGOScientific:2017qsa}
	{\bfseries Virgo, LIGO Scientific} Collaboration, B.~P. Abbott {\em et~al.},
	``{GW170817: Observation of Gravitational Waves from a Binary Neutron Star
		Inspiral},'' \href{http://dx.doi.org/10.1103/PhysRevLett.119.161101}{{\em
			Phys. Rev. Lett.} {\bfseries 119} no.~16, (2017) 161101},
	\href{http://arxiv.org/abs/1710.05832}{{\ttfamily arXiv:1710.05832 [gr-qc]}}.
	
	\bibitem{Abbott:2017gyy}
	{\bfseries Virgo, LIGO Scientific} Collaboration, B.~P. Abbott {\em et~al.},
	``{GW170608: Observation of a 19-solar-mass Binary Black Hole Coalescence},''
	\href{http://arxiv.org/abs/1711.05578}{{\ttfamily arXiv:1711.05578
			[astro-ph.HE]}}.
	
	\bibitem{Will:2005sn}
	C.~M. Will, ``{Post-Newtonian gravitational radiation and equations of motion
		via direct integration of the relaxed Einstein equations. III. Radiation
		reaction for binary systems with spinning bodies},''
	\href{http://dx.doi.org/10.1103/PhysRevD.71.084027}{{\em Phys. Rev.}
		{\bfseries D71} (2005) 084027},
	\href{http://arxiv.org/abs/gr-qc/0502039}{{\ttfamily arXiv:gr-qc/0502039}}.
	
	\bibitem{Blanchet:2013haa}
	L.~Blanchet, ``{Gravitational Radiation from Post-Newtonian Sources and
		Inspiralling Compact Binaries},''
	\href{http://dx.doi.org/10.12942/lrr-2014-2}{{\em Living Rev. Rel.}
		{\bfseries 17} (2014) 2},
	\href{http://arxiv.org/abs/1310.1528}{{\ttfamily arXiv:1310.1528 [gr-qc]}}.
	
	\bibitem{Jaranowski:2013lca}
	P.~Jaranowski and G.~Sch\"{a}fer, ``{Dimensional regularization of local
		singularities in the 4th post-Newtonian two-point-mass Hamiltonian},''
	\href{http://dx.doi.org/10.1103/PhysRevD.87.081503}{{\em Phys. Rev.}
		{\bfseries D87} (2013) 081503},
	\href{http://arxiv.org/abs/1303.3225}{{\ttfamily arXiv:1303.3225 [gr-qc]}}.
	
	\bibitem{Poisson:2011nh}
	E.~Poisson, A.~Pound, and I.~Vega, ``{The Motion of point particles in curved
		spacetime},'' {\em Living Rev.Rel.} {\bfseries 14} (2011) 7,
	\href{http://arxiv.org/abs/1102.0529}{{\ttfamily arXiv:1102.0529 [gr-qc]}}.
	
	\bibitem{Wald:2009ue}
	R.~M. Wald, ``{Introduction to Gravitational Self-Force},'' {\em Fundam. Theor.
		Phys.} {\bfseries 162} (2011) 253--262,
	\href{http://arxiv.org/abs/0907.0412}{{\ttfamily arXiv:0907.0412 [gr-qc]}}.
	[,253(2009)].
	
	\bibitem{Damour08}
	T.~Damour, A.~Nagar, M.~Hannam, S.~Husa, and B.~Brugmann, ``Accurate
	effective-one-body waveforms of inspiralling and coalescing black-hole
	binaries.'' Arxiv:0803.3162 [gr-qc], 2008.
	
	\bibitem{Kerr59a}
	R.~P. Kerr, ``The lorentz-covariant approximation method in general
	relativity,'' {\em Nuovo Cim.} {\bfseries XIII} no.~3, (1959) 469--491.
	
	\bibitem{Kerr59b}
	R.~P. Kerr, ``On the lorentz-covariant approximation method in general
	relativity. ii second approximation,'' {\em Nuovo Cim.} {\bfseries XIII}
	no.~3, (1959) 492--502.
	
	\bibitem{Kerr59c}
	R.~P. Kerr, ``On the lorentz-invariant approximation method in general
	relativity. iii the einstein-maxwell field,'' {\em Nuovo Cim.} {\bfseries
		XIII} no.~4, (1959) 673--689.
	
	\bibitem{Kerr60}
	R.~P. Kerr, ``On the quasi-static approximation in general relativity. iii the
	einstein-maxwell field,'' {\em Nuovo Cim.} {\bfseries XVI} no.~1, (1960)
	26--60.
	
	\bibitem{Gallo:2011tf}
	E.~Gallo and O.~M. Moreschi, ``{New derivation for the equations of motion for
		particles in electromagnetism},''
	\href{http://dx.doi.org/10.1103/PhysRevD.85.065005}{{\em Phys.Rev.}
		{\bfseries D85} (2012) 065005},
	\href{http://arxiv.org/abs/1112.5344}{{\ttfamily arXiv:1112.5344 [gr-qc]}}.
	
	\bibitem{Moreschi87}
	O.~M. Moreschi, ``General future asymptotically flat spacetimes,'' {\em Class.
		Quantum Grav.} {\bfseries 4} (1987) 1063--1084.
	
	\bibitem{Geroch87}
	R.~Geroch and J.~Traschen, ``Strings and other distributional sources in
	general relativity,'' {\em Phys. Rev. D} {\bfseries 36} (1987) 1017--1031.
	
	\bibitem{Szabados04}
	L.~B. Szabados, ``Quasi-local energy-momentum and angular momentum in general
	relativity,'' {\em Living Rev. Relativity} {\bfseries 12} ((2009) 4) .
	http://www.livingreviews.org/lrr-2009-4.
	
	\bibitem{Dixon:1970zza}
	W.~Dixon, ``{Dynamics of extended bodies in general relativity. I. Momentum and
		angular momentum},''
	{\em Proc.Roy.Soc.Lond.} {\bfseries A314} (1970) 499--527.
	
	\bibitem{Mathisson37}
	M.~Mathisson, ``New mechanics of material systems,'' {\em Gen.Rel.Grav.}
	{\bfseries 42} (2010) 1011--1048. Acta Phys. Pol. 6, 163 (1937).
	
	\bibitem{Mathisson40}
	M.~Mathisson, ``The variational equation of relativistic dynamics,'' {\em
		Proc.Camb.Philos.Soc.} {\bfseries 36} (1940) 331--350.
	
	\bibitem{Gallo:2016hpy}
	E.~Gallo and O.~M. Moreschi, ``{Constructing balanced equations of motion for
		particles in general relativistic theories: the general case},''
	\href{http://arxiv.org/abs/1609.02110}{{\ttfamily arXiv:1609.02110 [gr-qc]}}.
	
	\bibitem{Moreschi86}
	O.~M. Moreschi, ``On angular momentum at future null infinity,'' {\em Class.
		Quantum Grav.} {\bfseries 3} (1986) 503--525.
	
	\bibitem{Sachs62}
	R.~Sachs, ``Asymptotic symmetries in gravitational theory,'' {\em Phys. Rev.}
	{\bfseries 128} (1962) 2851--2864.
	
	\bibitem{Moreschi88}
	O.~M. Moreschi, ``{Supercenter of Mass System at Future Null Infinity},''
	\href{http://dx.doi.org/10.1088/0264-9381/5/3/004}{{\em Class.Quant.Grav.}
		{\bfseries 5} (1988) 423--435}.
	
	\bibitem{Moreschi98}
	O.~M. Moreschi and S.~Dain, ``Rest frame system for asymptotically flat
	space-times,'' {\em J. Math. Phys.} {\bfseries 39} no.~12, (1998) 6631--6650.
	
	\bibitem{Dain00'}
	S.~Dain and O.~M. Moreschi, ``General existence proof for rest frame system in
	asymptotically flat space-time,'' {\em Class. Quantum Grav.} {\bfseries 17}
	(2000) 3663--3672.
	
	\bibitem{Moreschi04}
	O.~M. Moreschi, ``Intrinsic angular momentum and center of mass in general
	relativity,'' {\em Class.Quantum Grav.} {\bfseries 21} (2004) 5409--5425.
	
	\bibitem{Gallo:2014jda}
	E.~Gallo and O.~M. Moreschi, ``{Intrinsic angular momentum for radiating
		spacetimes which agrees with the Komar integral in the axisymmetric case},''
	{\em Phys.Rev.} {\bfseries D89} (2014) 084009,
	\href{http://arxiv.org/abs/1404.2475}{{\ttfamily arXiv:1404.2475 [gr-qc]}}.
	
	\bibitem{Geroch73}
	R.~Geroch, A.~Held, and R.~Penrose, ``A space-time calculus based on pairs of
	null directions,'' {\em J. Math. Phys.} {\bfseries 14} (1973) 874--881.
	
	\bibitem{Penrose63}
	R.~Penrose, ``{Asymptotic properties of fields and space-times},''
	\href{http://dx.doi.org/10.1103/PhysRevLett.10.66}{{\em Phys.Rev.Lett.}
		{\bfseries 10} (1963) 66--68}.
	
	\bibitem{Moreschi95}
	L.Lehner and O.~Moreschi, ``On the definition of the center of mass for a
	system of relativistic particles,''
	\href{http://dx.doi.org/10.1063/1.530967}{{\em J. Math. Phys.} {\bfseries 36}
		no.~7, (1995) 3377--3394}.
	
	\bibitem{Moreschi88b}
	O.~M. Moreschi, ``On the limitation of approximation methods around stationary
	backgrounds for isolated systems,'' {\em Class. Quantum Grav.} {\bfseries 5}
	(1988) L53--L57.
	
	\bibitem{Walker:1979zk}
	M.~Walker and C.~M. Will, ``{Relativistic Kepler problem 2. Asymptotic behavior
		of the field in the infinite past},''
	\href{http://dx.doi.org/10.1103/PhysRevD.19.3495,
		10.1103/PhysRevD.20.3437}{{\em Phys. Rev.} {\bfseries D19} (1979) 3495}.
	[Erratum: Phys. Rev.D20,3437(1979)].
	
	\bibitem{Bel:1981}
	N.~D. J. I. J.~M. Luis~Bel, Thibaut~Damour, ``{Poincar\'e-invariant
		gravitational field and equations of motion of two pointlike objects: The
		postlinear approximation of general relativity},''
	\href{http://dx.doi.org/10.1007/BF00756073}{{\em Gen Relat Gravit.} {\bfseries
			13} (1981) 963}.
	
	\bibitem{Havas:1965}
	S.~F. Smith and P.~Havas, ``Effects of gravitational radiation reaction in the
	general relativistic two-body problem by a lorentz-invariant approximation
	method,'' \href{http://dx.doi.org/10.1103/PhysRev.138.B495}{{\em Phys. Rev.}
		{\bfseries 138} (Apr, 1965) B495--B508}.
	\url{http://link.aps.org/doi/10.1103/PhysRev.138.B495}.
	
	\bibitem{Rosemblum:1983}
	A.~Rosenblum, ``Effects of gravitational radiation reaction in the general
	relativistic two-body problem by a lorentz-invariant approximation method,''
	\href{http://dx.doi.org/http://dx.doi.org/10.1088/0305-4470/16/12/021}{{\em
			J. Phys. A} {\bfseries 16} (1983) 2575}.
	
	\bibitem{Gallo:2018a}
	E.~Gallo and O.~M. Moreschi, ``{Constructing balanced equations of motion for
		particles in general relativity: the null gauge case},''
	\href{http://arxiv.org/abs/18xx.yyyyy}{{\ttfamily arXiv:18xx.yyyyy [gr-qc]}}.
	
	\bibitem{Gallo:2017yys}
	E.~Gallo and O.~M. Moreschi, ``{Constructing balanced equations of motion for
		particles in general relativity: the harmonic gauge case},''
	\href{http://arxiv.org/abs/1711.08501}{{\ttfamily arXiv:1711.08501 [gr-qc]}}.
	
\end{thebibliography}

\end{document}